\documentclass[letterpaper,12pt]{article}
\usepackage{a4wide}
\usepackage[english]{babel}
\usepackage{amsmath,amssymb,amsthm,amsfonts,color,graphicx,epsfig,subfigure}

\title{Futures pricing in electricity markets based on stable CARMA spot models}
\author{Fred Espen Benth
\thanks{
Center of Mathematics for Applications (CMA),
University of Oslo,
P.O. Box 1053, Blindern,
N-0316 Oslo Norway
and
Department of Economics,
University of Agder,
Serviceboks 422,
N-4604 Kristiansand, Norway,
email: fredb@math.uio.no, http://folk.uio.no/fredb}
\and
Claudia Kl\"uppelberg
\thanks{Center for Mathematical Sciences, and Institute for Advanced Study, Technische Universit\"at M\"unchen, Boltzmannstrasse 3,  D-85748 Garching, Germany,
email: cklu@ma.tum.de, http://www-m4.ma.tum.de}
\and
Gernot M\"uller
\thanks{Center for Mathematical Sciences, Technische Universit\"at M\"unchen, Boltzmannstrasse 3,  D-85748 Garching, Germany,
email: mueller@ma.tum.de, http://www-m4.ma.tum.de/pers/mueller}
\and
Linda Vos
\thanks{Center of Mathematics for Applications (CMA),
University of Oslo,
P.O. Box 1053, Blindern,
N-0316 Oslo, Norway
and
Department of Economics,
University of Agder,
Serviceboks 422,
N-4604 Kristiansand, Norway,
email: linda.vos@cma.uio.no, http://folk.uio.no/lindavos}
}

\renewcommand{\baselinestretch}{1.2}

\numberwithin{equation}{section}

\newtheorem{theorem}{Theorem}[section]
\newtheorem{lemma}[theorem]{Lemma}
\newtheorem{remark}[theorem]{Remark}
\newtheorem{assumption}[theorem]{Assumptions}
\newtheorem{example}[theorem]{Example}
\newtheorem{proposition}[theorem]{Proposition}
\newtheorem{definition}[theorem]{Definition}
\newtheorem{corollary}[theorem]{Corollary}

\newtheorem{fig}[theorem]{Figure}

\newcommand{\bthe}{\begin{theorem}}
\newcommand{\ethe}{\end{theorem}}

\newcommand{\ble}{\begin{lemma}}
\newcommand{\ele}{\end{lemma}}

\newcommand{\bde}{\begin{definition}}
\newcommand{\ede}{\end{definition}}

\newcommand{\bco}{\begin{corollary}}
\newcommand{\eco}{\end{corollary}}

\newcommand{\bpr}{\begin{proposition}}
\newcommand{\epr}{\end{proposition}}

\newcommand{\brem}{\begin{remark}\rm}
\newcommand{\erem}{\halmos\end{remark}}

\newcommand{\bproof}{\begin{proof}}
\newcommand{\eproof}{\end{proof}}

\newcommand{\bexam}{\begin{example}\rm}
\newcommand{\eexam}{\halmos\end{example}}

\newcommand{\bfi}{\begin{fig}}
\newcommand{\efi}{\end{fig}}

\newcommand{\btab}{\begin{tab}}
\newcommand{\etab}{\end{tab}}

\newcommand{\ben}{\begin{enumerate}}
\newcommand{\een}{\end{enumerate}}

\newcommand{\beao}{\begin{eqnarray*}}
\newcommand{\eeao}{\end{eqnarray*}\noindent}

\newcommand{\beam}{\begin{eqnarray}}
\newcommand{\eeam}{\end{eqnarray}\noindent}

\newcommand{\barr}{\begin{array}}
\newcommand{\earr}{\end{array}}

\newcommand{\bdis}{\begin{displaymath}}
\newcommand{\edis}{\end{displaymath}\noindent}

\newcommand{\beq}{\begin{equation}}
\newcommand{\eeq}{\end{equation}}

\def\bbe{{\Bbb E}}
\def\bbr{{\Bbb R}}
\def\bbn{{\Bbb N}}

\def\CARMA{{\rm CARMA}}

\newcommand{\al}{{\alpha}}

\newcommand{\La}{{\Lambda}}
\newcommand{\ga}{{\gamma}}

\newcommand{\eps}{\varepsilon}

\newcommand{\wh}{\widehat}
\newcommand{\wt}{\widetilde}

\newcommand{\halmos}{\quad\hfill\mbox{$\Box$}}  

\def\e{{\textbf{e}}}
\def\bX{{\textbf{X}}}

\begin{document}


\begin{center}
{\LARGE \bf Futures pricing in electricity markets \\[2mm] based on stable CARMA spot models} \\[11mm]
{\large Fred Espen Benth\footnote{
Center of Mathematics for Applications (CMA),
University of Oslo,
P.O. Box 1053, Blindern,
N-0316 Oslo Norway
and
Department of Economics,
University of Agder,
Serviceboks 422,
N-4604 Kristiansand, Norway,
email: fredb@math.uio.no, http://folk.uio.no/fredb},
Claudia Kl\"uppelberg\footnote{
Center for Mathematical Sciences, and Institute for Advanced Study, Technische Universit\"at M\"unchen, Boltzmannstrasse 3,  D-85748 Garching, Germany,
email: cklu@ma.tum.de, http://www-m4.ma.tum.de},
Gernot M\"uller\footnote{
Center for Mathematical Sciences, Technische Universit\"at M\"unchen, Boltzmannstrasse 3,  D-85748 Garching, Germany,
email: mueller@ma.tum.de, http://www-m4.ma.tum.de/pers/mueller},
Linda Vos\footnote{
Center of Mathematics for Applications (CMA),
University of Oslo,
P.O. Box 1053, Blindern,
N-0316 Oslo, Norway
and
Department of Economics,
University of Agder,
Serviceboks 422,
N-4604 Kristiansand, Norway,
email: linda.vos@cma.uio.no, http://folk.uio.no/lindavos}
}
\end{center}
\vspace*{2mm}
\begin{abstract}
\small
\renewcommand{\baselinestretch}{1.0}
\small
We present a new model for the electricity spot price dynamics, which is able to capture seasonality, low-frequency dynamics and the extreme spikes in the market. 
Instead of the usual purely deterministic trend we introduce a non-stationary independent increments process for the low-frequency dynamics, 
and model the large fluctuations by a non-Gaussian stable CARMA process.
The model allows for analytic futures prices, and we apply these to model and estimate the whole market consistently. Besides standard parameter estimation, 
an estimation procedure is suggested, where we fit the non-stationary trend using futures data with long time until delivery, and  a robust $L^1$-filter to 
find the states of the CARMA process. The procedure also involves the empirical and theoretical risk premiums which -- as a by-product -- are also estimated.
We apply this procedure to data from the German electricity exchange EEX, where we split the empirical analysis into base load and peak load prices. 
We find an overall negative risk premium for the base load futures contracts, except for contracts close to delivery, where a small positive risk premium is detected. 
The peak load contracts, on the other hand, show a clear positive risk premium, when they are close to delivery, while the contracts in the longer end also have a negative premium.
\end{abstract}
\vspace*{2mm}

\noindent
\begin{tabbing}
\> secondary:\,\,\,60G10, 60G51, 91B70
\end{tabbing}

\vspace{2mm}


\vspace{1mm}

\newpage

\renewcommand{\baselinestretch}{1.5}
\normalsize

\section{Introduction}\label{s1}

In the last decades the power markets have been liberalized world-wide, and there is a large
interest for modelling power spot prices and derivatives. Electricty spot prices are
known to be seasonally varying and mean-reverting.
Moreover, a distinctive characteristic of spot prices is the large spikes that
occur due to sudden imbalances in supply and demand, for example, when a large production utility
experiences a black-out or temperatures are suddenly dropping. Typically in these markets, different
production technologies have big variations in costs, leading to a very steep
supply curve. Another characteristic of electricity is the lack of efficient
storage possibilities. Many spot price models have been suggested for electricity, 
and we refer to Eydeland and Wolyniec~\cite{EW} and Benth,
\v{S}altyt\.{e} Benth and Koekebakker~\cite{BBK} for a discussion on various models 
and other aspects of modelling of energy markets.

In this paper we propose a two-factor aritmethic spot price model with
seasonality, which is analytically feasible
for pricing electricity forward and futures contracts. The spot price model consists of a
continuous-time autoregressive moving average factor driven by a stable L\'evy process
for modelling the stationary short-term  variations, and a non-stationary long-term factor given by a
L\'evy process. We derive futures prices under a given pricing measure, and propose to
fit the spot model by a novel optimization algorithm using spot and futures price data
simultaneously. We apply our model and estimation procedure on price data observed at the German 
electricity exchange EEX.

In a seminal paper by Schwartz and Smith~\cite{SS} a two-factor model for commodity spot prices
is proposed. Their idea is
to model the short-term logarithmic spot price variations as a mean-reverting Ornstein-Uhlenbeck
process driven by a Brownian motion, reflecting the drive in prices towards its equilibrium
level due to changes in supply and demand. But, as argued by Schwartz and Smith~\cite{SS},
there may be significant uncertainty in the
equilibrium level caused by inflation, technological innovations, scarceness of fuel resources
like gas and coal etc.. To account for such long-term randomness in prices, Schwartz and
Smith~\cite{SS} include a second non-stationary factor being a drifted Brownian motion,
possibly correlated with the short-term variations. They apply their model to crude oil futures
traded at NYMEX,
where the non-stationary part is estimated from futures prices, which are far from delivery. Mean-reversion
will kill off the short-term effects from the spot on such futures, and they can thus be applied
to filter out the non-stationary factor of the spot prices.

This two-factor model is applied to electricity prices by Lucia and Schwartz~\cite{LS}.
Among other models, they fit an arithmetic two-factor model with deterministic seasonality
to electricity spot prices collected from the Nordic electricity exchange NordPool. 
Using forward and futures prices they fit the model, where the distance
between theoretical and observed prices are minimized in a least squares sense.

A major critiscism against the two-factor model considered in Lucia and Schwartz~\cite{LS} is
the failure to capture spikes in the power spot dynamics. By using Brownian motion driven
factors, one cannot explain the sudden large price spikes frequently observed in spot data.
Multi-factor models, where one or more factors are driven by jump processes, may mend this.
For example, Benth, Kallsen and Meyer-Brandis~\cite{BKMB} suggest an arithmetic spot model
where normal and spike variations in the prices are separated into different factors
driven by pure-jump processes. In this way one may
model large price increases followed by fast speed of mean-reversion together with
a ``base component''-behaviour, where price fluctuations are more slowly varying
around a mean level. Such multi-factor models
allow for analytic pricing of forward and futures contracts.

A very attractive alternative to these multi-factor models are given by the class of
continuous-time autoregressive moving-average processes, also called CARMA processes.
These processes incorporate in an efficient way memory effects and mean-reversion, and
generalize Ornstein-Uhlenbeck processes in a natural way (see Brockwell \cite{Bro2001}).
As it turns out, (C)ARMA processes fit power spot prices extremely well, as demonstrated by
Bernhardt, Kl\"uppelberg and Meyer-Brandis~\cite{BKM} and Garcia, Kl\"uppelberg and
M\"uller~\cite{GKM}. In \cite{BKM} an ARMA process with stable innovations, and in \cite{GKM} 
a CARMA(2,1)-model driven by a stable L\'evy
process are suggested and empirically studied on power spot price data collected from the
Singapore and German EEX markets, respectively.
A CARMA(2,1) process may be viewed on a discrete-time scale as an autoregressive process of order 2, 
with a moving average order 1.
By invoking a stable L\'evy
process to drive the CARMA model, one is blending spikes and small innovations in prices into
one process. We remark in passing that a CARMA dynamics
has been applied to model crude oil prices at NYMEX by Paschke and Prokopczuk~\cite{PP} and
interest-rates by Zakamouline, Benth and Koekebakker~\cite{BKZ}.

We propose a generalization of the stable CARMA model of Garcia et al.~\cite{GKM} by including a
long-term non-stationary factor being a general L\'evy process. The model allows for
analytical pricing of electricity futures, based on pricing measures, which preserve the
L\'evy property of the driving processes. More precisely, we apply an Esscher measure transform
to the non-stationary part, and a transform which maps the stationary stable process into a tempered stable. 
Due to the semi-affine structure of the model, the futures and forward prices becomes explicitly
dependent on the states of the CARMA model and the non-stationary factor.

The CARMA-based factor in the spot model accounts for the short-term variations in
prices and will be chosen stationary. By a CARMA model with a higher order autoregressive part we may include
different mean-reversion speeds, such that we can mimic the behaviour of a multi-factor model
accounting for spikes and base variations separately. The moving average part is necessary to model 
the observed dependence structure. The stable L\'evy process may have
very big jumps, which then can explain spike behaviour in the prices. The smaller variations
of the stable L\'evy process model the base signal in power prices. As it turns out from
our empirical investigations using market price data from the EEX, the non-stationary long-term
behaviour may accurately be modelled using a normal inverse Gaussian (NIG) L\'evy process.
We filter out the non-stationary part from observing futures prices, which are far from
delivery. The influence of the stationary CARMA factor is then not present, and the data
shows a significant non-normal behaviour. This is in contrast to the choice suggested by
Schwartz and Smith~\cite{SS} and applied by Lucia and Schwartz~\cite{LS}. Moreover,
we find that a CARMA(2,1) model is accurately explaining the mean-reversion and
memory effects in the spot data.

A novelty of our paper apart from the generalizing existing one and two factor models,
is our
estimation procedure. Lucia and Schwartz~\cite{LS} propose an iterative algorithm
for estimating their two-factor model to NordPool electricity data, where they minimize the least-squares
distance between the theoretical and observed forward and futures prices to find the
risk-neutral parameters. In order to find the theoretical prices, they must have the states of
the two factors in the spot model accessible. Since these are not directly observable, they
choose an iterative scheme, where they start with a guess on the parameter values, find the
states minimizing distance, update parameters by estimation, find the states minimizing the
distance etc. until convergence is reached. We propose a different approach, utilizing the
idea in Schwartz and Smith~\cite{SS} that the non-stationary factor is directly observable, at least
approximately, from forward prices, which are far from delivery. We apply this to filter out
the non-stationary factor. The  CARMA-part is then observable from the spot prices, where
seasonality and the non-stationary term is subtracted. Since we work with stable processes, 
which do not have finite second moments, $L^2$-filters can not be used
to find the states of the CARMA-process. We propose a simple $L^1$-filter being more
robust with respect to spikes in spot data to do this.
The problem we are facing is to determine, what contracts to use for filtering out
the non-stationary part. To find an optimal ``time-to-maturity'' which is sufficiently
far from delivery, so that the futures prices behave as the non-stationary factor and at the same time
provide a sufficiently rich set of data, we use an optimization algorithm, which minimizes the
least square distance between the empirical and theoretical risk premium. In order to
find the risk premia, we must have all the parameters of the model available, which in turn
can only be found, if we know which futures contracts can be used for filtering the long-term factor. 
We implement an algorithm, which estimates all model parameters for futures contracts withs different 
times to maturities and minimizes the distance to the empricial risk premium. 

We apply our model and estimation scheme to data from the German EEX (European Energy Exchange)
market where we use spot prices as well as futures prices of contracts with a delivery period
of one month. Our empirical studies cover both base load and peak load contracts, 
where base load contracts are settled against the average of all hourly spot prices 
in the delivery period. Peak load futures contracts are settled against 
the average of hourly spot price in peak periods of the delivery period. 
The peak load period is the period between 8 a.m. and 8 p.m., during every working day.
As a first summary, we can say that the results for both base and peak load data
are in general rather similar. However, the peak load data show a more extreme
behaviour. The average risk premium decays when time to maturity increases, 
and is negative for contracts in the longer end of the futures curve.
This points towards a futures market, where producers use the contracts 
for hedging and in return accept to pay a premium to insure their
production, in accordance with the theory of normal backwardation. 
The risk premium is completely determined by the effect of the long-term factor, 
which induces a close to linear decay as a function of ``time-to-maturity''.
We see that for the base load contracts the risk premium in the short end
of the curve is only slightly positive. The risk premium
is negative for contracts starting to deliver in about two months or later.
On the other hand, the peak load contracts have a clear positive risk premium, 
which turns to a negative one for contracts with delivery in
about four months or later. The positive risk premium
for contracts close to delivery tells us that the demand side 
(retailers and consumers) of the market is willing to
pay a premium for locking in electricity prices as a hedge against spike risk 
(see Geman and Vasicek~\cite{GV}).

Our results are presented as follows.
In Section~\ref{s2} we present the two-factor spot model, and we compute analytical
futures prices along with a discussion of pricing measures in Section~3. Section~4 explains
in detail the estimation steps and the procedure applied to fit the model to data. The results of this estimation procedure applied to EEX data is presented and discussed in Section~5. We conclude in Section~6.

Throughout we use the following notation. For a matrix $D$ we denote by $D^*$ its transposed, and $I$ is the identity matrix. For $p\in\bbn$ we denote by $\e_p$ the $p$-th unit vector. The matrix exponential $e^{At}$ is defined by its Taylor expansion $e^{At} = I + \sum_{n=1}^\infty  \frac{(At)^n}{n!} $ with identity matrix $I$. We also denote by $\log^+ x=\max(\log x,0)$ for $x\in\mathbb{R}$.

\section{The spot price dynamics}\label{s2}

In most electricity markets, like the EEX, hourly spot prices for the delivery of 1~MW of electricity are quoted. As is usual in the literature on electricity spot price modeling, one assumes a continuous-time model and estimates it on the discretely observed {\it daily} average spot prices. We refer, for instance, to Lucia and Schwartz~\cite{LS} and Benth, \v{S}altyt\.{e} Benth and Koekebakker~\cite{BBK} for more details.

We generalize the $\alpha$-stable (C)ARMA model of Bernhardt, Kl\"uppelberg and Meyer-Brandis~\cite{BKM} and Garcia, Kl\"uppelberg and M\"uller~\cite{GKM} by adding a non-stationary stochastic component in the trend of the spot dynamics.
By modeling the trend as a combination of a stochastic process and a deterministic seasonality function rather than only a deterministic seasonality function, which seems common in most models, we are able to describe the low frequent variations of the spot dynamics quite precisely.  As it turns out, this trend will explain a significant part of the futures price variations and lead to an accurate estimation of the risk premium in the EEX market.

A two-factor spot price model for commodities, including a mean-reverting short-time dynamics and a non-stationary long-term variations component was first suggested by
Schwartz and Smith~\cite{SS}, and later applied to electricity markets by Lucia and Schwartz~\cite{LS}. Their models were based on Brownian motion driven stochastic processes, more precisely, the sum of an Ornstein-Uhlenbeck (OU) process with a drifted Brownian motion. We significantly extend this model to include jump processes and higher-order memory effects in the dynamics.

Let $(\Omega,\mathcal{F},\{\mathcal{F}_t\}_{t\ge0},P)$ be a complete filtered probability space satisfying the usual
conditions of completeness and right continuity.
We assume the spot price dynamics
\begin{equation}
\label{spot-model}
S(t) = \Lambda(t) + Z(t) + Y(t),\quad t\ge0,
\end{equation}
where $\Lambda$ is a deterministic trend/seasonality function and $Z$ is a L\'evy process with zero mean.  The process $Z$ models the low-frequency non-stationary dynamics of the spot, and can together with $\Lambda$ be interpreted as the long-term factor for the spot price evolution.
The process $Y$ accounts for the stationary short-term variations.
We will assume that $Y$ and $Z$ are independent processes. We follow Garcia et al.~\cite{GKM} and Bernhardt et al.~\cite{BKM} and suppose that $Y$ is a stationary CARMA-process driven by an $\alpha$-stable L\'evy process.

\subsection{The stable CARMA-process}

We introduce stationary CARMA($p,q$)-L\'evy processes (see Brockwell \cite{Bro2001}) and discuss its relevant properties.

\bde[CARMA($p,q$)-L\'evy process]\label{def:stablecarma}\hspace{0cm}\\
A \CARMA$(p,q)$-L\'evy process $\lbrace Y(t) \rbrace_{t \geq 0}$ (with $0 \leq q < p$) driven by a L\'evy-process $L$ is defined as the solution of the state space equations
\beam\label{eq;statespace}
Y(t) &=& \textbf{b}^* \textbf{X}(t)\\
d\textbf{X}(t) &=& A\textbf{X}(t) dt + \textbf{e}_p dL(t), \label{eq;statesp}
\eeam
 with
 \begin{align*}
  \textbf{b} = \left(
\begin{array}{c}
b_0\\
b_1\\
\vdots \\
b_{p-2}\\
b_{p-1}
\end{array}
\right),\quad
 \textbf{e}_p = \left(
\begin{array}{c}
0\\
0\\
\vdots \\
0\\
1
\end{array}
\right)
, \quad
A = \left(\begin{array}{ccccc}
0 &1 & 0& \cdots &0 \\
0&0&1& \cdots & 0\\
\vdots & \vdots & \vdots & \ddots & \vdots \\
0 & 0 & 0& \vdots & 1 \\
-a_p & -a_{p-1} & -a_{p-2} & \vdots & -a_1
\end{array}
\right).
\end{align*}
where $a_1,\ldots , a_p, b_0, \ldots, b_{p-1}$ are possibly complex-valued coefficients such that $b_q = 1$ and $b_j = 0$
for $q<j\le p$. For $p=1$ the matrix $A$ is to be understood as $A = -a_1$.
\ede

The driving process $L$ of $Y$ will be a non-Gaussian $\al$-stable L\'evy process $\{L(t)\}_{t\ge0}$ with characteristic function
given by $\ln\mathbb{E}e^{izL(t)}=t\phi_L(z)$ for $z\in\bbr$, where,
\begin{gather} \label{charf}
\phi_L(z) = \begin{cases}
-\gamma^\alpha|z|^\alpha(1-i\beta(\textnormal{sign } z) \tan\left(\frac{\pi \alpha}{2} \right) ) + i\mu z \quad \textnormal{ for } \alpha \not= 1, \\
-\gamma |z|(1+ i \beta \frac{2}{\pi} ( \textnormal{sign }z) \log |z| ) + i \mu z \quad \textnormal{ for } \alpha =1.
\end{cases}
\end{gather}
The sign function is defined by $\text{sign } z=-1$ for $z<0$, $\text{sign } z=1$ for $z>0$ and $\text{sign }0=0$, respectively.
Further, $\alpha\in(0,2)$ is the shape parameter, $\gamma>0$ the scale, $\beta\in[-1,1]$ the skewness, and $\mu$ the location parameter.
If $\gamma=1$ and $\mu=0$, then $L$ is called standardized.

The parameter $\alpha\in(0,2)$  determines the tail of the distribution function of $L(t)$ for all $t\ge0$.
Moreover, only moments strictly less than $\alpha$ are finite, so that no second moment exists.
This implies also that the autocorrelation function does not exist.
For further properties on stable processes and L\'evy processes, we refer to the monographs by Samorodnitsky and Taqqu~\cite{ST} and Sato~\cite{S}.

The solution of the SDE \eqref{eq;statesp} is a $p$-dimensional Ornstein-Uhlenbeck (OU)  process given by
\begin{align}\label{eq;X}
\textbf{X}(t) = e^{A(t-s)} \textbf{X}(s) + \int_s^t e^{A(t-u)} \textbf{e}_p dL(u), \quad 0\le s< t,
\end{align}
where the stable integral is defined as in Ch.~3 of Samorodnitsky and Taqqu~\cite{ST}.
From \eqref{eq;statespace} we find that $Y$ is given by
\begin{align}\label{eq;Y}
Y(t) = \textbf{b}^*e^{A(t-s)} \textbf{X}(s) + \int_s^t \textbf{b}^*e^{A(t-u)} \textbf{e}_p dL(u), \quad 0\leq s<t.
\end{align}
Equations \eqref{eq;statespace} and \eqref{eq;statesp} constitute the state-space representation of the formal $p$-th order SDE
\begin{gather}\label{formaleq}
a(D)Y(t) = b(D) DL(t), \quad t \geq 0,
\end{gather}
where $D$ denotes differentiation with respect to $t$, and
 \begin{eqnarray}
a(z) &:=& z^p + a_1 z^{p-1}  + \cdots + a_p \label{a-ch-polyn}\\
b(z) &:=& b_0 + b_1 z + \cdots + b_q z^q \label{b-ch-polyn}
\end{eqnarray}
are the characteristic polynomials.
Equation \eqref{formaleq} is a natural continuous-time analogue of the linear difference equations, which define an ARMA process (cf. Brockwell and Davis~\cite{Brock_book}).

Throughout we assume that $Y$ and $\bX$ are stationary in the sense that all finite dimensional distributions are shift-invariant.
Based on Proposition~2.2 of Garcia et al.~\cite{GKM} (which summarizes results by Brockwell and Lindner~\cite{BL}) we make the following assumptions to ensure this:
\begin{assumption}\label{assump} Stationarity of CARMA-process. \\
(i)\phantom{ii} The polynomials $a(\cdot)$ and $b(\cdot)$ defined in \eqref{a-ch-polyn} and \eqref{b-ch-polyn}, resp., have no common zeros. \\
(ii)\phantom{i} $\bbe \left[ \log^+ |L(1)| \right]<\infty$. \\
(iii)           All eigenvalues of $A$ are distinct and have strictly negative real parts. 
\end{assumption}
Assumption (ii) and (iii) imply that $\mathbf{X}$ is a causal $p$-dimensional OU process, hence also $Y$ is causal.

\brem
Our model is a significant generalization of the two-factor dynamics by Schwartz and Smith~\cite{SS} and Lucia and Schwartz~\cite{LS}. Among various models, Lucia and Schwartz~\cite{LS} suggested a two factor dynamics of the spot price evolution based on a short term Gaussian OU process and a long-term drifted Brownian motion.
In our framework a Gaussian OU process would correspond to a Gaussian CARMA(1,0) process.
It is clear that such a Gaussian model cannot capture the large fluctuations in the spot price, like for example spikes, and jump processes seem to be the natural extension. Based on the studies of Bernhardt et al.~\cite{BKM} and Garcia et al.~\cite{GKM}, $\alpha$-stable processes are particularly suitable for the short-term dynamics in the spot price evolution. Furthermore, empirical analysis of electricity spot price data from Singapore and Germany in \cite{BKM} and \cite{GKM} show strong statistical evidence for full CARMA processes to capture the dependency structure of the data.
As for the long-term baseline trend, we shall see in Section~\ref{sec;empirical} that a normal inverse Gaussian L\'evy process is preferable to a Gaussian process in a data study from the German electricity exchange EEX.
\erem

\subsection{Dimensionality of CARMA-processes}\label{s211}

A more standard model in electricity is to describe the spot by a sum of several OU processes, where some summands describe the spike behavior and others the baseline dynamics (see for example the model by Benth et al.~\cite{BKMB}).
A CARMA process is in a sense comparable to such models, as we now discuss.

By Assumption~\ref{assump}(iii) $A$ has full rank, i.e. it is diagonalizable with $e^{At} = U e^{D t} U^{-1}$.
Here, $D$ is a diagonal matrix with the eigenvalues $\lambda_1, \ldots, \lambda_p$ of $A$ on the diagonal and $U$ is the full rank matrix having the eigenvectors of $A$ as columns.
Since all eigenvalues have negative real parts, all components of $e^{At}$ are mean reverting.
Each component of the vector $e^{A(t-s)} \bX(s)$ from \eqref{eq;Y} will therefore mean revert at its own speed, where
the speed of mean reversion is a linear combination of the diagonal elements of $e^{D t}$.

As we shall see in a simulation example of a CARMA(2,1)-process in Section~\ref{Lfilter}, it captures the situation, where a first component has a slower rate of mean-reversion than the second (see Fig.~\ref{F;states}). This is similar to a two-factor spot model, where the base and spike components of the spot price evolution are separated into two OU processes with different speeds of mean reversion.
The advantage of working with a stable CARMA process, as we propose, is that it is possible to capture the distribution of the small and large jumps in one distribution. Since extreme spikes are rather infrequently observed, it
is difficult to calibrate the spike component in a conventional two-factor model; this has been observed in Kl\"uppelberg et al.~\cite{KMS}.
With our CARMA-model, we avoid the difficult question of spike identification and filtering.

\section{The futures price dynamics}\label{s3}

In commodity markets, futures contracts are commonly traded on exchanges, including electricity, gas, oil, and coal.
In this section we derive the futures price dynamics based on the $\alpha$-stable CARMA spot model \eqref{spot-model}.
Appealing to general arbitrage theory (see e.g.~Duffie \cite{duffie}, Ch.1), we define the futures price $f(t,\tau)$ at time $t$ for
a contract maturing at time $\tau$ by
\begin{gather}\label{eq;forward}
f(t,\tau)=\mathbb{E}_{Q}\left[S(\tau)\,|\,\mathcal{F}_t\right]\,, \quad 0 \leq t \leq \tau<\infty\,,
\end{gather}
where $Q$ is a risk neutral probability measure. This definition is valid as long as $S(\tau)\in L^1(Q)$.
In the electricity market, the spot cannot be traded, and every $Q\sim P$ will be a risk neutral probability (see Benth et al.~\cite{BBK}).
For example, $Q=P$ is a valid choice of a pricing measure.
In that case, the condition $S(\tau)\in L^1(P)$ is equivalent to a tail parameter $\alpha$ of the stable process $L$ being strictly larger than one, and a process  $Z$ with finite expectation. In real markets one expects a risk premium and hence it is natural to use a pricing measure
$Q\neq P$.  We will discuss possible choices of risk neutral probability measures $Q$ in Section~\ref{s4}.

Based on our spot price model, we find the following explicit dynamics of the futures price for a given class of risk neutral probability measures:

\bthe\label{forwarddynamic}
Let $S$ be the spot dynamics as in \eqref{spot-model}, and suppose that $Q\sim P$ is such that $L$ and $Z$ are L\'evy processes under $Q$. 
Moreover, assume that the processes $Z$ and $L$ have finite first moments under $Q$.
Then the futures price dynamics for $0\le t\le \tau$ is given by
\beao
f(t, \tau) &\!\!=\!\!&  \Lambda(\tau) + Z(t) +  \textbf{b}^*e^{A(\tau-t)} \textbf{X}(t) 
+ (\tau - t) \mathbb{E}_Q[Z(1)] + \textbf{b}^* A^{-1} \left( I-e^{A(\tau-t)} \right) {\bf e}_p \, \mathbb{E}_{Q}[L(1)]\,.
\eeao
\ethe

\begin{proof}
Using \eqref{eq;forward}, $f(t,\tau) = \mathbb{E}_Q[S(\tau)\mid\mathcal{F}_t] = \Lambda(\tau) + \mathbb{E}_Q[Z(\tau)\mid\mathcal{F}_t] 
+ \mathbb{E}_Q[Y(\tau)\mid\mathcal{F}_t]$.
Since $Z$ is a L\'evy process under $Q$, we find
\begin{align*}
\mathbb{E}_{Q}[Z(\tau)\mid\mathcal{F}_t] &= Z(t) + \mathbb{E}_{Q}[Z(\tau)-Z(t)\mid\mathcal{F}_t]
\, = \, Z(t) + (\tau - t) \mathbb{E}_Q[Z(1)]\,.
\end{align*}
Now denote by $M(u)=L(u)-\mathbb{E}_Q[L(1)]u$ for $t\le u\le\tau$, which has zero mean. 
Then, by partial integration,
\beao
\lefteqn{\mathbb{E}_{Q} \left[\int_t^\tau \textbf{b}^*e^{A(\tau-u)} \textbf{e}_p\, dM(u) \right]}\\
&=& \mathbb{E}_{Q}\left[\textbf{b}^*\textbf{e}_p M(\tau) - \textbf{b}^*e^{A(\tau-t)} \textbf{e}_p M(t)\right]
-  \int_t^\tau \textbf{b}^*A e^{A(\tau-u)} \textbf{e}_p\, \mathbb{E}_{Q}[M(u)] du = 0\,,
\eeao
$$
\mbox{\hspace*{-2.5cm}which implies} \qquad \mathbb{E}_{Q} \left[\int_t^\tau \textbf{b}^*e^{A(\tau-u)} \textbf{e}_p\, dL(u) \right] \, = \,\mathbb{E}_{Q}[L(1)] \, \int_t^\tau \textbf{b}^*e^{A(\tau-u)} \textbf{e}_p \, du \,.
$$
Hence, the CARMA part of the spot dynamics converts to
\beao
\lefteqn{\mathbb{E}_{Q}[Y(\tau) \mid \mathcal{F}_t] = 
\mathbb{E}_{Q} \left[ \textbf{b}^*e^{A(\tau-t)} \textbf{X}(t) + \int_t^\tau \textbf{b}^*e^{A(\tau-u)} \textbf{e}_p \, dL(u) \vert \mathcal{F}_t \right]} \\
& \!\!\!\!\!\!= \!\!\!&   \textbf{b}^*e^{A(\tau-t)} \textbf{X}(t)  \!+\!  \mathbb{E}_{Q} \left[\int_t^\tau \textbf{b}^*e^{A(\tau-u)} \textbf{e}_p \, dL(u) \right] 
 = \textbf{b}^*e^{A(\tau-t)} \textbf{X}(t)  \!+\!\!  \int_t^\tau \textbf{b}^* e^{A(\tau-u)} \textbf{e}_p \, du \, \mathbb{E}_{Q}[L(1)] \\
& \!\!\!\!\!\!= \!\!\!&   \textbf{b}^*e^{A(\tau-t)} \textbf{X}(t)  + \textbf{b}^* A^{-1} \left( I-e^{A(\tau-t)} \right){\bf e}_p \, \mathbb{E}_{Q}[L(1)]\,.
\eeao
Combining the terms yields the result.
\end{proof}

In electricity markets the futures contracts deliver the underlying commodity over a period rather than at a fixed maturity time $\tau$.
For instance, in the German electricity market contracts for delivery over a month, a quarter or a year, are traded.
These futures are sometimes referred to as swaps, since during the delivery period a fixed (futures) price of energy is swapped against a floating (uncertain) spot price. The futures price is quoted as the price of 1 MWh of power and, therefore, it is settled against the average spot price over the delivery period. Hence, the futures price $F(t,T_1,T_2)$ at time $0\le t\le T_1<T_2$ for a contract with delivery period
$[T_1,T_2]$ is defined as
\begin{align}\label{swapprice}
F(t,T_1,T_2) = \mathbb{E}_{Q} \left[\frac1{T_2-T_1}\int_{T_1}^{T_2} S(\tau) d\tau \Big{|} \mathcal{F}_t \right]\,,
\end{align}
where we have assumed that settlement of the contract takes place at the end of the delivery period, $T_2$.


Using Theorem~\ref{forwarddynamic} we derive by straightforward integration the swap price dynamics \\
$F(t,T_1,T_2)$ from \eqref{swapprice}.

\bco \label{forwardprice}
Suppose all assumptions of Theorem~\ref{forwarddynamic} are satisfied.
Then,
\begin{equation*}
F(t, T_1, T_2) = \frac{1}{T_2-T_1} \int_{T_1}^{T_2} \Lambda(\tau) d\tau + Z(t)+ \frac{{\bf b}^*A^{-1}}{T_2 - T_1} \left(e^{AT_2} -  e^{AT_1} \right) e^{-At} \textbf{X}(t) \label{swapstates}+\Gamma_Q(t,T_1,T_2)
\end{equation*}
where
\begin{align*}
\Gamma_Q(t,T_1,T_2)&=\left(\frac{1}{2}(T_2+T_1)-t\right) \mathbb{E}_Q[Z(1)]-\frac{{\bf b}^*A^{-2}}{T_2-T_1} \left( e^{AT_2}- e^{AT_1} \right) e^{-At} {\bf e}_p \, \mathbb{E}_{Q}[L(1)]  \\
&\qquad + \textbf{b}^* A^{-1} {\bf e}_p\, \mathbb{E}_{Q}[L(1)] \,.
\end{align*}
\eco
\begin{proof}
By the Fubini Theorem, we find 
$$
F(t,T_1,T_2)=\frac1{T_2-T_1}\int_{T_1}^{T_2}f(t,\tau)\,d\tau\,.
$$
Applying Theorem~\ref{forwarddynamic} and integrating yield the desired result.
\end{proof}
The risk premium is defined as the difference between the futures price and the predicted spot, that is, in terms of electricity futures contracts,
\begin{equation}
\label{def-riskpremium}
R_{pr}(t,T_1,T_2)=F(t,T_1,T_2)-\mathbb{E}\left[\frac{1}{T_2-T_1}\int_{T_1}^{T_2}S(\tau)\,d\tau\,|\,\mathcal{F}_t\right]\,.
\end{equation}
From Cor.~\ref{forwardprice} we find that the theoretical risk premium for a given pricing measure $Q$ is
\beam
\lefteqn{R(t,T_1,T_2) \, = \, \Gamma_Q(t,T_1,T_2)-\Gamma_P(t,T_1,T_2) \nonumber}\\
&=&\left(\frac{1}{2}(T_2+T_1)-t\right) \mathbb{E}_Q[Z(1)]-\frac{{\bf b}^*A^{-2}}{T_2-T_1} \left( e^{AT_2}- e^{AT_1} \right) e^{-At} {\bf e}_p \, \left(\mathbb{E}_{Q}[L(1)]-\mathbb{E}[L(1)]\right)  \nonumber\\
& &\qquad + \textbf{b}^* A^{-1} {\bf e}_p\, \left(\mathbb{E}_{Q}[L(1)]-\mathbb{E}[L(1)]\right) 
\label{riskpremium}\,.
\eeam
Here, we used the assumption that $Z$ has zero mean under $P$. The first term gives a trend in "time to maturity" implied by the non-stationarity part $Z$ in the spot price dynamics. "Time to maturity" is here interpreted as the time left to the middle of the delivery period. The two last terms are risk premia contributions from the CARMA short-term spot dynamics. They involve an explicit dependence on the speeds of mean-reversion of the autoregressive parts and the memory in the moving-average part. We will apply the risk premium in the empirical analysis of spot and futures data from the EEX.

\subsection{Equivalent measure transforms for L\'evy and $\alpha$-stable processes}\label{s4}


In this subsection we discuss a class of pricing measures that will be used for the specification of the futures price dynamics.

We require from the pricing measure that $Z$ and $L$ preserve their L\'evy property and the independence.
For this purpose, we consider probability measures $Q=Q_L\times Q_Z$, where $Q_L$ and $Q_Z$ are measure changes for $L$ and $Z$, respectively (leaving the other process unchanged). Provided $Z$ has exponential moments, a standard choice of measure change is given by the Esscher transform (see Benth et al.~\cite{BBK}, Section~4.1.1). Note that $L$, the $\alpha$-stable process in the CARMA-dynamics, does not have exponential moments.

We define the density process of the Radon-Nikodym derivative of $Q_Z$ as
\begin{equation}
\frac{dQ_Z}{dP}\Big|_{\mathcal{F}_t}=\exp\left(\theta_Z Z(t)-\phi_Z(\theta_Z)t\right),\quad t\ge0,
\end{equation}
for a constant $\theta_Z\in\bbr$ and $\phi_Z$ being the log-moment generating function of $Z(1)$ (sometimes called the
cumulant function of $Z$). 
In order to make this density process well-defined,
exponential integrability of the process $Z$ up to the order of $\theta_Z$ must be assumed.
Under this change of measure, the L\'evy measure of $Z$ will be exponentially tilted by $\theta_Z$, that is, if we denote the L\'evy measure of $Z$ (under $P$) by $\nu(dx)$, then its L\'evy measure under $Q_Z$ becomes $\nu_{Q_Z}(dx)=\exp(\theta_Z)\nu(dx)$
(see Benth et al.~\cite{BBK}, Section.~4.1.1-4.1.2 for details).

To choose a risk neutral measure $Q_L$ is a more delicate task. We know from Sato~\cite{S}, Theorems~33.1 and~33.2, that equivalent measures $Q$ exists for stable processes, however, it seems difficult to construct one which preserves the stable property.
As an alternative, one may introduce the class of {\it tempered stable processes} (see e.g. Cont and Tankov~\cite{C&T}, Chapter 9), and apply
standard Esscher transformation on these.

A tempered stable process is a pure jump L\'evy process, where the stable-like behavior is preserved for the small jumps.
However, the tails are tempered and, therefore, extreme spikes are less likely to be modeled with the tempered stable process.
The L\'evy measure is given by
\beam\label{tstable}
\nu_{TS} (dx) &=& \frac{c_+ e^{\theta_L x}}{x^{1+\alpha}}\, 1_{(0,\infty)}(x)\, dx + \frac{c_- e^{\theta_L |x|}}{|x|^{1+\alpha}}\, 1_{(-\infty,0)}(x)\, dx\,.
\eeam
Here, $\theta_L\leq 0$ and $c_-,c_+ \in \mathbb{R}_+$.
A consequence of the tempering is that certain exponential moments exist. Tempering of a stable distribution results in a {\it tempered} stable distribution, and is analogous to taking an Esscher transform of the stable process
using a negative parameter $\theta_L$ on the positive jumps, and a positive parameter $-\theta_L$ on the negative jumps.

In particular,  define $q: \mathbb{R} \mapsto \mathbb{R}$ as
$
q(x) :=   e^{\theta_L x} \, 1_{(0,\infty)}(x) + e^{\theta_L |x|}\,1_{(-\infty,0)}(x)
$
for some constant $\theta_L < 0$.  
Suppose the stable distribution $L$ has (under $P$) the characteristic triplet $(\ga_L,0,\nu_L)$, where
\begin{align*}
\nu_L(dx) = \frac{c_+}{x^{1+\alpha}}\, 1_{(0,\infty)}(x)\, dx + \frac{c_-}{|x|^{1+\alpha}}\, 1_{(-\infty,0)}(x)\, dx
\end{align*}
is the L\'evy measure of our stable process $L$. The parameters $c_+, c_-$ can be matched to the parameters in \eqref{charf}, using Example 2.3.3 of Samorodnitsky and Taqqu \cite{ST}. Then the tempered stable measure $Q_L$, with characteristic triplet $(\ga_{TS}, 0, \nu_{TS})$ is {\it equivalent} to the physical probability measure $P$ (see Cont and Tankov~\cite{C&T}, Proposition~9.8), with drift parameter $\gamma_{TS}$ is given by
\begin{align}\label{meanTS}
\ga_{TS}= \begin{cases}
\ga_L &  0<\alpha <1 \\
\ga_L+  \int_{\{|x|<1\}} x(q(x) -1) \nu_L(dx) &  1 <\alpha <2
\end{cases}
\end{align}
and the L\'evy measure $\nu_{TS}$ is given by
$
\nu_{TS} (dx) = q(x) \nu_L(dx).
$
The special case of a Cauchy distribution ($\alpha = 1$) is left out since one is not able to define the L\'evy-Kintchine formula using a truncation of the small jumps and the large jumps given by $1_{|x|>1}$.
For our applications it is of particular value to know the expectations of $L(1)$ under $P$ and $Q_L$.

\ble\label{riskpremium0}
Let $L$ be an $\al$-stable L\'evy process under $P$ with $\al\in(1,2)$.
Find $Q_L$ by stable tempering for $\theta_L<0$ as in \eqref{tstable}.
Then the difference in mean of $L(1)$ under $Q_L$ and $P$ is given by
\begin{equation}\label{riskpremiumrepr}
\mathbb{E}_{Q_L}[L(1)] - \mathbb{E}[L(1)] = \Gamma (1- \alpha)(-\theta_L)^{\alpha-1} \left( c_+  - c_- \right)\,,
\end{equation}
where $\Gamma$ is the gamma function.
\ele

\bproof
Using \eqref{meanTS} and the L\'evy-Khintchine formula (e.g. Cont and Tankov \cite{C&T}, Prop.~3.13) for $ 1< \alpha <2$ we obtain
\begin{align} \label{EQPL}
\mathbb{E}[L(1)] - \mathbb{E}_{Q_L}[L(1)] &=  \ga_L - \ga_{TS} + \int_{\{|x| >1\}} x (\nu_L - \nu_{TS})(dx) \nonumber \\
&=  \ga_L - \ga_{TS} + \int_{\{|x| >1\}} x(1-q(x)) \nu_L(dx)\nonumber \\
&= c_-\int_{-\infty}^0 (1-e^{\theta_L|x|})\frac{dx}{x^{1+\alpha}}
+ c_+\int_0^\infty (1-e^{\theta_L x})\frac{dx}{x^{1+\alpha}}  \nonumber \\
&= -\Gamma (1- \alpha)(-\theta_L)^{\alpha-1} \left( c_+  - c_- \right)\,,
\end{align}
where we have used partial integration on the two integrals and l'Hospital's rule to obtain the last identity.
This proves the result.
\eproof

\brem\label{R;riskpremiumrepr} By altering $\theta_L$ one can match any relevant mean change in the risk premium $\mathbb{E}_Q[L(1)]- \mathbb{E}[L(1)]$, as long as this can be obtained by a negative choice of $\theta_L$. This turns out to be appropriate for our applications.
\erem

\section{Fitting the model to German electricity data}\label{sec;empirical}

Our data are daily spot and futures prices from July 1, 2002 to June 30, 2006 
(available from {\sl http://eex.com}). We fitted our model both to base load and peak load data, respectively. 
The futures contracts considered in this analysis are the Phelix-Base-Month-Futures and the Phelix-Peak-Month-Futures.
{\em Base load futures contracts} are settled against the average of all hourly spot prices in the delivery period. 
{\em Peak load futures contracts}, on the other hand, are settled against the average of the hourly spot prices 
in peak periods of the delivery period. The peak period is counted as the hours between 8 a.m. and 8 p.m. every working day 
during the delivery period. 
The time series of daily spot prices used for our combined statistical analysis is taken to match the futures contracts:
for the base load contracts we use the full time series consisting of daily observations including weekends
(i.e., we have 7 observations per week), while in the case of peak load contracts the weekends are excluded
(i.e., we have 5 observations per week).

\begin{figure}[ht]
\begin{center}
\includegraphics[width = 14cm]{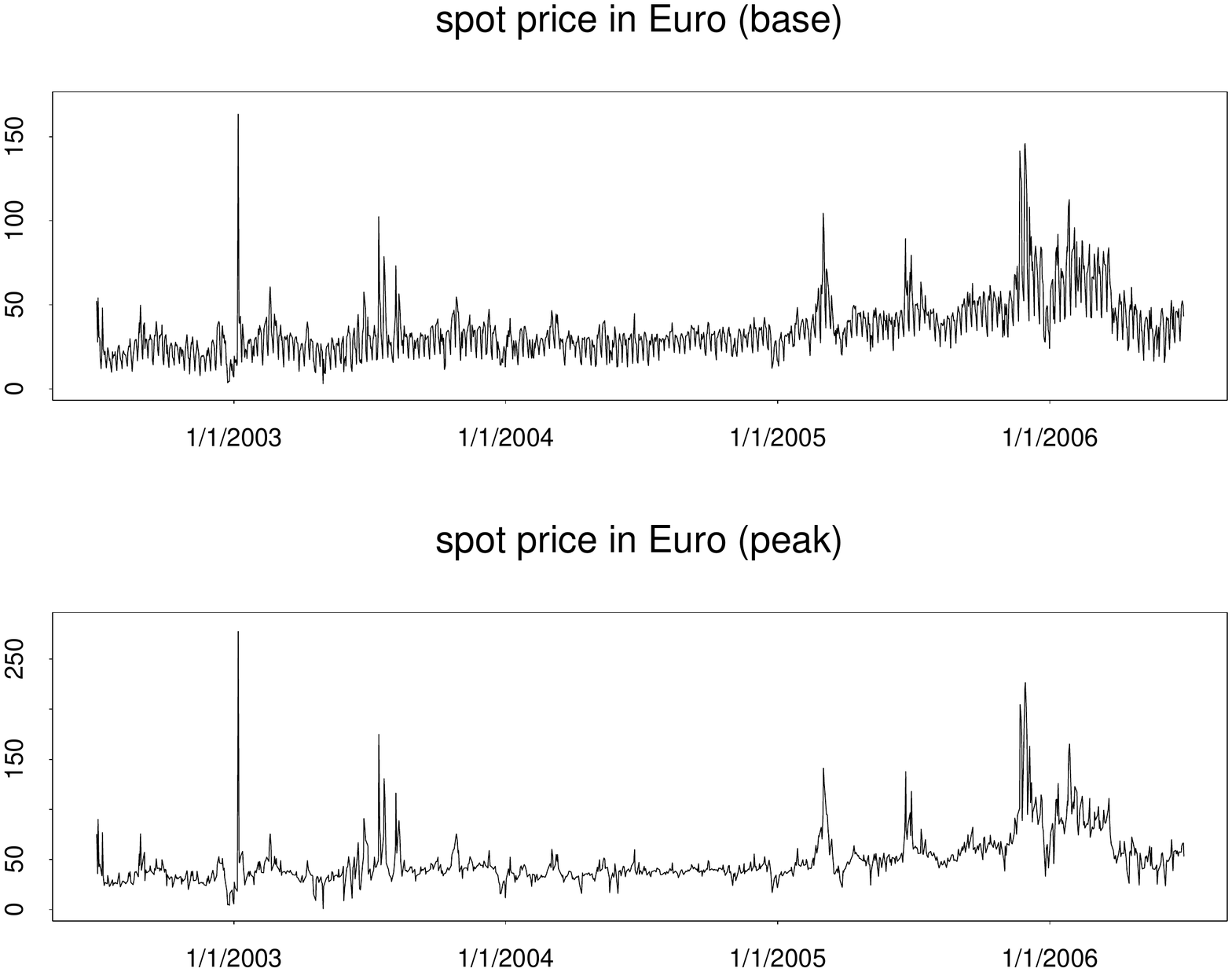}
\caption{{\it Daily spot prices from July 1, 2002 to June 30, 2006, base load (top) and peak load (bottom).  
}}\label{PLOTspotpricesBP}
\end{center}
\end{figure}

\begin{figure}[ht]
\begin{center}
\includegraphics[width = 14cm]{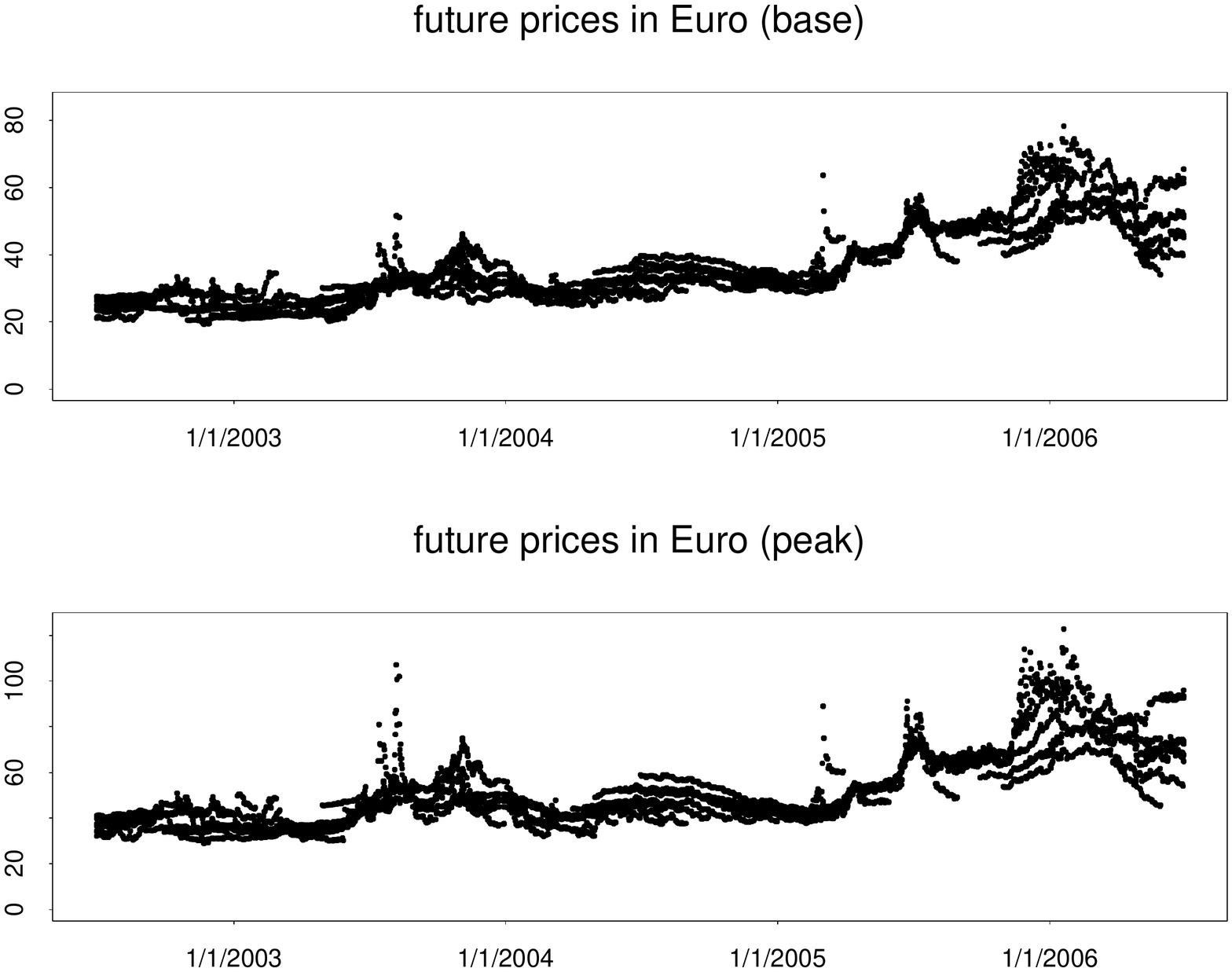}
\caption{{\it Daily futures prices from July 1, 2002 to June 30, 2006, base load (top) and peak load (bottom).
}}\label{PLOTfutpricesBP}
\end{center}
\end{figure}

Figures \ref{PLOTspotpricesBP} and \ref{PLOTfutpricesBP} show the spot and futures prices for both base load and peak load.
From these plots we can see similar patterns of the base and peak load data, however, peak load data are more extreme. 
Note that all plots cover the same time period; however, for the base spot data we have 1461 observations, 
whereas for the peak spot data we have only 1045 observations in the same period, due to the missing weekends.

The estimation procedure for our model consists of several steps, 
which are explained in the following.

\subsection{Seasonality function $\Lambda$}
\label{est;Lambda}
The estimation of the deterministic trend component $\Lambda$ is a delicate question. A mis-specification of the trend has a 
significant effect on the subsequent analysis, in particular, on the risk premium.
Motivated by the seasonality functions used in Bernhardt et al.~\cite{BKM} and Garcia et al.~\cite{GKM},
we take the seasonality function of the peak load contracts as a combination of a linear trend and some periodic function
\beam\label{season-peak}
\Lambda_p(t) &=& c_1 + c_2 t + c_3 \cos\left(\frac{2\pi t}{261}\right) + c_4 \sin \left(\frac{2\pi t}{261}\right).
\eeam

Note that we choose a slightly simpler seasonality function than  Bernhardt et al.~\cite{BKM} and Garcia et al.~\cite{GKM}, 
only taking the mean level, a linear trend and a yearly periodicity (modeling the weather difference between summer and winter) 
into account. Weekly periodicity in peak load contracts is not that pronounced, since weekends are not considered 
in peak load data. Since no new trading information is entering during the weekend (trading takes place during weekdays), 
we will adjust the periodicity to 261 and consider the peak load contracts as a continuous process on all non-weekend days.

In the base load prices a clear weekly seasonality is visible. Weekend prices are in general lower than during the rest
of the week, and over the week one observes the pattern that Monday and Friday have prices lower than in the middle of the week. 
Therefore we include a weekly term in the base load seasonality function:
\beam\label{season-base}
\Lambda_b(t) &=& c_1 + c_2 t + c_3 \cos\left(\frac{2\pi t}{365}\right) + c_4 \sin \left(\frac{2\pi t}{365}\right)+ c_5 \cos\left(\frac{2\pi t}{7}\right) + c_6 \sin \left(\frac{2\pi t}{7}\right).
\eeam
Since time $t$ is running through the weekends, a yearly periodicity of 365 is chosen.

\begin{table}
\begin{center}
\begin{tabular}{|l|c|c|c|c|c|c|}
\hline
&$\mathbf{c_1}$&$\mathbf{c_2}$& $\mathbf{c_3}$&$\mathbf{c_4}$&$\mathbf{c_5}$&$\mathbf{c_6}$\\
\hline
{\bf Base}&$ 19.4859 $&$ 0.0217 $&$ -2.8588 $&$ 0.6386 $&$ -6.7867 $&$ 2.8051 $\\
\hline
{\bf Peak}&$ 30.7642 $&$ 0.0349 $&$ -2.5748 $&$ 1.5762 $&$         $&$        $\\
\hline
\end{tabular}\caption{{\it Estimated parameters of the seasonality function $\Lambda(\cdot)$.}}\label{T;season}
\end{center}.
\end{table}

In the following, we will analyse both data sets, base load data and peak load data (spot and futures, respectively).
For simplicity we will suppress the indices $p$ for peak load and $b$ for base load.

\begin{figure}[ht]
\begin{center}
\includegraphics[width = 14cm]{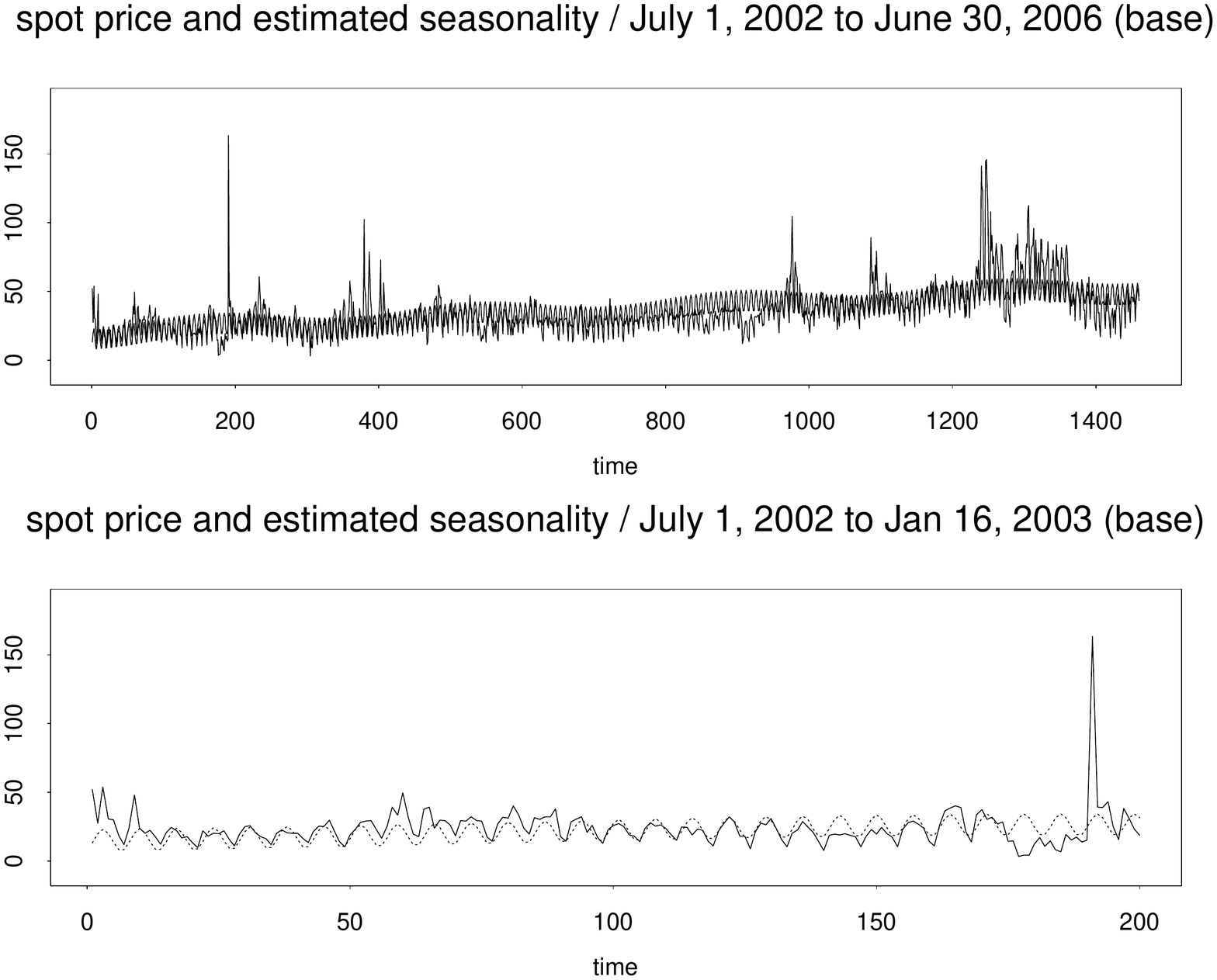}
\caption{{\it Base spot prices and estimated seasonality function. 
Top: whole period (1461 observations). 
Bottom: first 200 observations.
}}\label{SpotBASEtrend}
\end{center}
\end{figure}

This seasonality functions can be estimated using a robust least-squares estimate on the data; 
see Table~\ref{T;season} for the resulting estimates. 
The first plot of Figure \ref{SpotBASEtrend} shows the 1461 observations of the base data set 
together with the estimated trend and seasonality curve, the second plot zooms into this plot and shows 
these curves for the first 200 observations. From this second plot we can clearly see that  
the daily seasonality is caputered by $\Lambda$ quite accurately. For the peak data over the same period
(consisting of 1045 observations) we get similar plots (hence, we
omit them here), except of the fact that $\Lambda$ does not show a weekly seasonality in this case.
The overall growth rate was very small during the data period, justifying a linear term as a first order approximation. 
Note that we have introduced the non-stationary stochastic process $Z$ to absorb all stochastic small term effects in the seasonality. This term will play a prominent role for the futures prices later.

Subtracting the estimated seasonality function from the spot data leaves us with the 
reduced model $Z(\cdot)+Y(\cdot)$, where we have neglected in our notation the fact 
that we have subtracted only an estimator of $\La(\cdot)$.

Next we want to estimate both components $Z$ and $Y$ invoking the deseasonalised spot price data and the futures prices.
We will exploit the fact that the futures prices far from delivery will have a dynamics approximately given by 
the non-stationary trend component $Z$. Only relatively close to delivery, large fluctuations in the spot price
dynamics are reflected in the futures prices. Since it is not clear how far away from delivery we need to be 
before the approximation of futures prices by $Z$ works well, we will invoke an optimization routine to find the 
optimal distance. For this purpose, we introduce the notation $u:= \frac{1}{2}(T_1+T_2) - t$, which will be referred to as
{\em ``time to maturity''}.

Denote by $\hat{u}^*$ the optimal time to maturity (we will define what we understand by ``optimal'' below), where 
futures contracts with time to maturity $u\geq \hat{u}^*$ have a dynamics approximately behaving like the non-stationary term.
How big to choose $\hat{u}^*$ is not possible to determine {\it a priori}, since we must analyse the error in an 
asymptotic consideration of the futures prices (see \eqref{eq;1} and \eqref{eq;2} below). This error is highly 
dependent on the parameters in the spot price model, which we do not yet know.
In the end, $\hat{u}^*$ should be chosen so that the error in the risk premium 
estimation is minimal; cf.~Section~\ref{estRisk}.

The estimation will, therefore, be repeated using different values of $u^*$ (this parameter 
will sometimes be called {\em threshold} in the following); i.e., we choose a subset 
$U^* := \left[ u^*_{\mbox{\footnotesize min}}, u^*_{\mbox{\footnotesize max}} \right] 
\subseteq \left[ v/2, M_f \right]$,
where $v$ is the average delivery period and $M_f$ is the maximal time to maturity observed in the futures data set,
and perform the steps of the Sections~\ref{estZ}--\ref{estRisk} below repeatedly for $u^* \in U^*$. 
For each value $u^* \in U^*$ the error in the risk premium
is calculated, and $\hat{u}^*$ is the value which minimizes this error among all $u^* \in U^*$. 
This {\em optimal} threshold $\hat{u}^*$ is then considered as {\em final} choice of $u^*$ for the calculation of all estimates including the processes $Z$ and $Y$ and for the CARMA parameters.
 
One should keep in mind that for too large $u^*$ there is only few data available with $u\geq u^*$, which
yields unreliable estimates. 
Since we count the time to maturity as number of trading days until the mid
of the delivery period (which has length $v$), time to maturity is always at least $v/2$; hence we do not consider any $u^*$ smaller than $v/2$.
Overall, we decided to choose
$u^*_{\mbox{\footnotesize min}}=\lceil v/2 \rceil$ 
(which is 16 for the base load and 11 for the peak load data) 
and $u^*_{\mbox{\footnotesize max}}=M_f/2$ (note that $M_f$ is 200 days for the base load and 144 days for the peak 
load contracts). 
As we will see later, the optimal $\hat{u}^*$ in our data examples is
quite small, so that this choice of $u^*_{\mbox{\footnotesize max}}$ is completely satisfying.


Next we want to explain in detail, how we separate $Z$ and $Y$ for a given fixed time to maturity.
Consequently, we perform the model estimation for all $\frac{1}{2}(T_1+T_2)\le u^*\le 200(146)$ and take all futures prices for the estimation procedure, whose time to maturity  $u\ge u^*$.

We explain each step in the estimation procedure in detail:

\subsection{Filtering the realization of the non-stationary stochastic process $Z$}\label{estZ}

Recall the futures price $F(t,T_1,T_2)$ in Corollary~\ref{forwardprice}.
Since we assume that the high-frequency CARMA term $Y$ is stationary, it holds for fixed length of delivery $T_2-T_1$ that
\begin{align}
\lim_{T_1,T_2\rightarrow\infty}\frac{{\bf b}^*A^{-1}}{T_2 - T_1} \left(e^{AT_2} -  e^{AT_1} \right) e^{-At} \textbf{X}(t) &= 0 \label{eq;1}\\
\lim_{T_1,T_2\rightarrow\infty} \frac{{\bf b}^*A^{-2}}{T_2-T_1} \left( e^{AT_2}- e^{AT_1} \right) e^{-At} {\bf e}_p \, \mathbb{E}_{Q}[L(1)] &=0\,.  \label{eq;2}
\end{align}
Hence, in the long end of the futures market, the contribution from $Y$ to the futures prices may be considered as negligible.
In particular, from the futures price dynamics (Corollary~\ref{forwardprice}) we find for $[T_1,T_2]$ far into the future (that is, $t$ much smaller than $T_1$) that
\begin{align}\label{eq;F}
\widetilde{F}(t,T_1,T_2) &:= F(t,T_1,T_2) - \frac{1}{T_2-T_1} \int_{T_1}^{T_2} \Lambda(\tau) d\tau \nonumber \\
&  \approx Z(t) + {\bf b}^* A^{-1} {\bf e}_p \, \mathbb{E}_Q[L(1)] + \left(\frac12(T_1+T_2)-t\right)\, \mathbb{E}_Q[Z(1)].
\end{align}
Recalling the notation $u := \frac{1}{2} (T_1+T_2) - t$ coined ``time-to-maturity'', we slightly abuse the notation and introduce $\widetilde{F}(t,u):=\widetilde{F}(t,T_1,T_2)$.

For $u\ge u^*$, we approximate
\begin{align}\label{eq;risk}
\mu_{\widetilde{F}}(u) &:= \mathbb{E}[\widetilde{F}(t,u)] \nonumber \\
&  \approx \mathbb{E}[Z(t)] + {\bf b}^* A^{-1} {\bf e}_p \, \mathbb{E}_Q[L(1)] + u\, \mathbb{E}_Q[Z(1)]  \nonumber \\
&=  {\bf b}^* A^{-1} {\bf e}_p \, \mathbb{E}_Q[L(1)]+  u \, \mathbb{E}_Q[Z(1)]  \nonumber \\
& =: \, C + u \,\mathbb{E}_Q[Z(1)]\,,
\end{align}
where we have used the zero-mean assumption of $Z$ under $P$.
This approximative identity can now be used for a robust linear regression on the time to maturity $u$, 
in order to estimate the real numbers $C$ and $\mathbb{E}_Q[Z(1)]$.
Knowing these two parameters enables us to filter out the realization of the process $Z$.
According to Equation \eqref{eq;F} we obtain
\begin{align}\label{eq;z}
\wh{Z}(t) = \wh  Z \left(\frac12(T_1+T_2)-u\right) = \frac{1}{{\rm card }\,U(t,u^*)} \sum_{(u,T_1,T_2) \in U(t,u^*)} 
\left[ \widetilde{F}(t,T_1,T_2) - \wh C - u\, \wh{\mathbb{E}}_Q[Z(1)] \right],
\end{align}
where $U(t,u^*) := \left\{ (u,T_1,T_2) \in \bbr^3 \,|\, u \ge u^* \mbox{ and } \exists F(t,T_1,T_2): \frac12(T_1+T_2)-t=u \right\}$.
\brem\label{R;Z}
Note that after estimating the CARMA parameters, we can also find an estimate
for $\bbe_{Q}[L(1)]$ simply by taking $\wh\bbe_{Q}[L(1)]= \wh C (\wh{\bf b}^* \wh A^{-1} {\bf e}_p )^{-1}$.
\erem

\brem\label{R;Z2}
We recall that the futures market at EEX is not open for trade during the weekend. 
Therefore, using our estimation procedure, we do not get any observations of $Z$ during weekends. 
We will assume that $Z$ is constant and equal to the Friday value over the weekend, when filtering 
the non-stationary part of the spot in the base load model. 
One may argue that this strategy could lead to large observed jump of $Z$ on Monday morning, when all 
information accumulated over the weekend is subsumed at once. We will return to this question in 
Section \ref{disZ}.
\erem

\subsection{Estimation of the CARMA parameters}\label{estCAR}
\begin{figure}[ht]
\begin{center}
\includegraphics[width = 14cm]{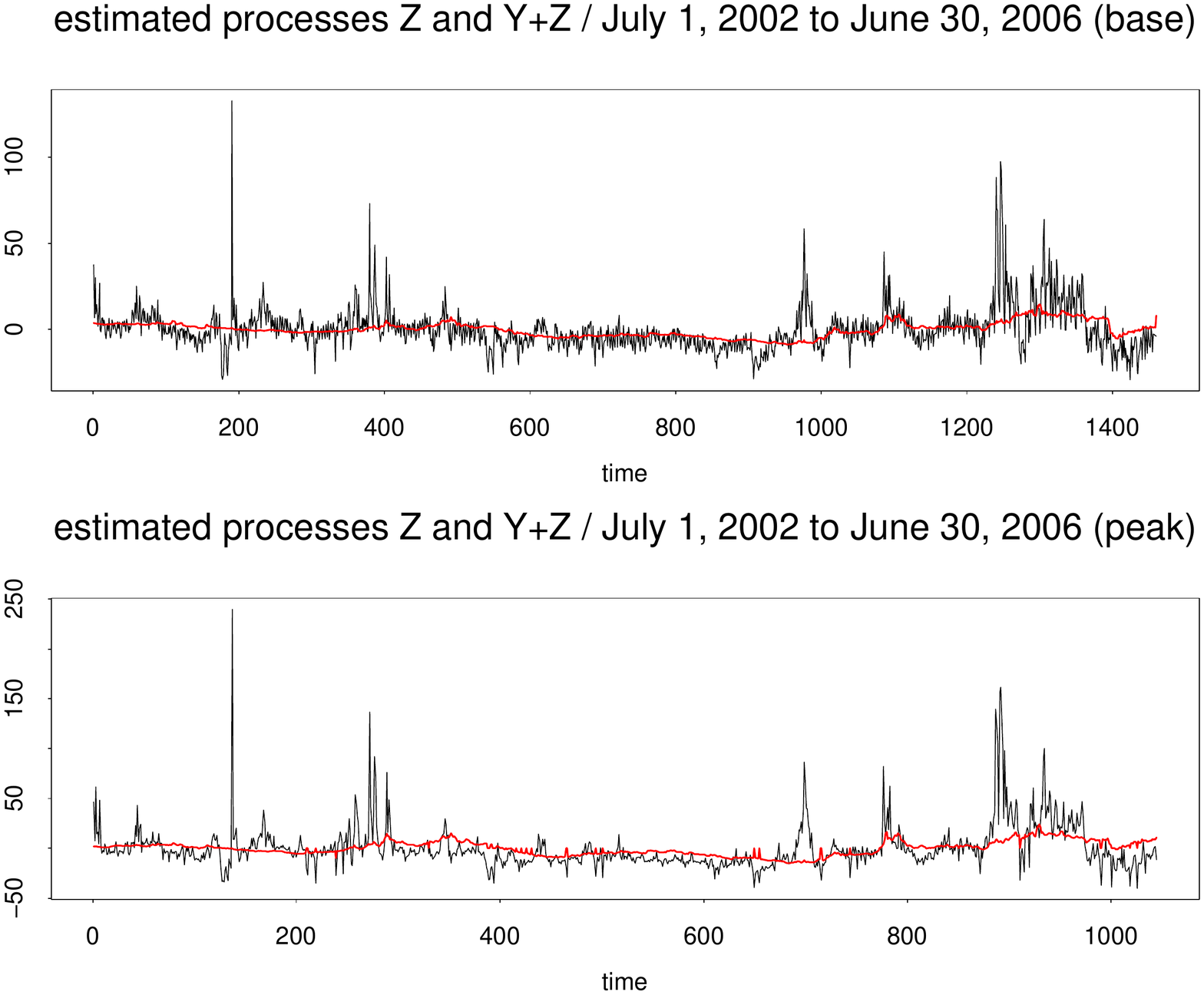}
\caption{{\it Estimated processes $Z$ (red) and $Y+Z$ (black) in the period July 1, 2002, to June 30, 2006,
for both the base load data (top) and the peak load data (bottom).
}}\label{PLOTYandZ}
\end{center}
\end{figure}

Recall our spot-model \eqref{spot-model}
\begin{align*}
S(t) = \Lambda(t) + Y(t) + Z(t),\quad t\ge0.
\end{align*}
After $\Lambda(\cdot)$ and $Z(\cdot)$ have been estimated in Subsections \ref{est;Lambda} and \ref{estZ}, respectively, 
a realization of the CARMA-process $Y$ can be found by subtracting both from the spot price.
Figure \ref{PLOTYandZ} shows the estimated processes $Z$ (dotted red) and $Y+Z$ (black) for both the base and the peak load data, for
the full period July 1, 2002 to June 30, 2006 and for $u^*=16$ exemplarily, each. Obviously, 
the process $Z$ captures the medium-range fluctuations and
$Y$ the short-range fluctuations of the detrended and deseasonalized process $Y+Z$. 

Again we keep in mind that the process $Y$ is the result of some estimation procedure. There exists a number of papers devoted to the estimation of the CARMA parameters in $L^2$ (see for instance Brockwell et al.~\cite{BDY}, Tsai and Chan \cite{TC}).
Methods can be based either directly on the continuous-time process or on a discretised version.
The latter relates the continuous-time dynamics to a discrete time ARMA process.
The advantage of this method is obvious, since standard packages for the estimation of ARMA processes may be used in order to estimate the parameters of the corresponding CARMA process.
Some care, however, is needed since this approach does not work in all cases. Brockwell and collaborators devote several papers to the embedding of ARMA processes in a CARMA process; cf. \cite{Emb,Emb2}. Not every ARMA$(p,q)$ process is embeddable in a CARMA$(p,q)$ process.

From Assumptions~\ref{assump}, it follows by Proposition 2 of Brockwell et al.~\cite{BDY} (cf. also Garcia et al.~\cite{GKM}, Prop.~2.5) that every CARMA$(p,q)$  process $Y$ observed at discrete times can be represented as an autoregressive process of order $p$ with a more complex structure than a moving average process for the noise.
For such a discretely observed CARMA process $Y$ on a grid with grid size $h$, denoting the sequence of observations by $\{y_n\}_{n\in\bbn}$; i.e. $y_n \triangleq Y(nh)$, Prop.~3 of Brockwell et al.~\cite{BDY} gives
\begin{align}\label{eq;ARMACARMA}
\prod_{i=1}^p(1-e^{\lambda_i}B) y_{n} = \eps_n.
\end{align}
Here, $B$ is the usual backshift operator and $\{\eps_n\}_{n\in\bbn}$ is the noise process, which has representation
\begin{align}\label{eq;noise}
\eps_n = \sum_{i=1}^p \kappa_i \, \prod_{j \not= i} (1-e^{\lambda_j h} B)  \int_{(n-1)h}^{nh} e^{\lambda_i (nh -u)} dL(u)\,.
\end{align}
The constants $\kappa_i$ are given by $\kappa_i := b(\lambda_i)/a'(\lambda_i)$ and $\lambda_1,\ldots,\lambda_p$ are the eigenvalues of $A$.
The process $\{\eps_n\}_{n\in\bbn}$ is $p$-dependent.
When $L$ has finite variance, $\eps_n$ has a moving average representation; cf. Brockwell et al.~\cite{BDY}, Proposition~3.2.1. However, for the case of infinite variance this is no longer true and causes problems.

Davis~\cite{D}, Davis and Resnick~\cite{DR} and Mikosch~et al.~\cite{MGKA} have proved that ordinary $L^2$-based estimation methods for ARMA parameters may be used for $\alpha$-stable ARMA processes, although they have no finite second moments.
Moreover, Davis and Resnick~\cite{DR} showed that the empirical autocorrelation function of an $\al$-stable ARMA process yields a consistent estimator of the linear filter of the model, although the autocorrelation function of the process does not exist.
Hence the parameters $a_1,\ldots,a_p$ can be consistently estimated by $L^2$-methods.
In \cite{DR} it has also been shown that the rate of convergence is faster than in the $L^2$-case.

Already in the analysis performed in Garcia et al.~\cite{GKM} the CARMA(2,1) process has been found to be optimal. Although our model is slightly different, it turns out that this CARMA dynamics is still preferrable for $Y$ based on the AIC model selection criterion.
Hence, in the following example we spell out the above equations for the case of a stable CARMA(2,1) model.

\begin{example}\label{stable1}\rm [The CARMA(2,1) process]\\
By applying \eqref{eq;ARMACARMA} and \eqref{eq;noise} for the case of a CARMA(2,1) process, we find the discrete-time representation for a gridsize $h>0$,
\begin{align*}
y_n = \left(e^{\lambda_1 h} + e^{\lambda_2h} \right) y_{n-1} - e^{(\lambda_1+\lambda_2)h}\,y_{n-2} + \eps_n,
\end{align*}
where $\eps_n$ is given by
\begin{align*}
\eps_n &=  \int_{(n-1)h}^{nh} \left(\kappa_1 \, e^{\lambda_1 (nh-u)} + \kappa_2\, e^{\lambda_2(nh -u)}\right) dL(u) \\
& \quad +  \int_{(n-2)h}^{(n-1)h} \left(\kappa_1 \, e^{\lambda_2 h } \,e^{\lambda_1 (nh-u)} + \kappa_2\, e^{\lambda_1 h} \,e^{\lambda_2(nh -u)}\right) dL(u)\,.
\end{align*}
The two integrals in the noise are independent. It is, however, not possible to recover the noise by simple multiplication and subtraction as in the ARMA case. The actual relation of two successive noise terms $\eps_n$ and $\eps_{n+1}$ is based on the continuous realization of $\{L(t)\}_{t\ge0}$ in the relevant intervals, which is unobservable.
\halmos
\end{example}

For the mapping of the estimated ARMA parameters to the corresponding CARMA parameters we observe that equation \eqref{eq;ARMACARMA} is a complex way to express that $\{e^{-\lambda_i h}\}_{i=1,\ldots,p}$ are the roots of the autoregressive polynomial $\phi(z) = 1 - \phi_1 z - \cdots -\phi_p z^p$ of the ARMA process.
We proceed, therefore, as follows for identifying the CARMA parameters from the estimated ARMA process:
\begin{itemize}
\item Estimate the coefficients $\phi_1, \ldots, \phi_p$ of the ARMA process
\item Determine the distinct roots $\xi_i$ for $i=1,\ldots,p$ of the characteristic polynomial.
\item Set $\lambda_i = -\log(\xi_i)/h$, where we recall that $h$ denotes the grid size.
\end{itemize}
Because of the simple structure of the autoregressive matrix $A$ of the CARMA process we can calculate the characteristic polynomial $P$ of the matrix A as
\begin{align*}
P(\lambda) =  (-1)^p \left(\lambda^{p} + a_1 \lambda^{p-1} + \cdots + a_p \right).
\end{align*}
Since the $\lambda_i$ are the eigenvalues of $A$, we know that $P(\lambda_i) = 0$ for $i=1,\ldots,p$. Hence, given the eigenvalues $\lambda_i$ of the matrix $A$ we recover the coefficients $a_1,\ldots,a_p$ by solving a system of $p$ linear equations.

We estimate the moving average parameters based on the autocorrelation function. For its estimation we apply a least absolute deviation algorithm based on the empirical and theoretical autocorrelation functions of the CARMA process.
The theoretical autocorrelation function of $y$ takes the form
\begin{align*}
\gamma_y(s) = {\bf b}^*e^{A|s|} \Sigma {\bf b},\quad s>0,
\end{align*}
where the matrix $\Sigma$ is given by
\begin{align*}
\Sigma = \int_0^\infty e^{Au} {\bf e}_p {\bf e}_p^* e^{A^*u} du = -{\bf A}^{-1} {\bf e}_p {\bf e}_p^*\,.
\end{align*}
In this representation, ${\bf A}^{-1}$ is the inverse of the operator ${\bf A}: X\mapsto AX + XA^*$ and can be represented as $\mathrm{vec}^{-1} \circ ((A \otimes I_p) + (I_p \otimes A))^{-1} \circ \mathrm{vec}$ (see Pigorsch and Stelzer~\cite{PS}).
Using the above procedure for the estimation of the moving average parameter ${\bf b}$ is based on second order structure and, therefore, not straightforward to use for stable processes. In practice this procedure works and we can use it to estimate the moving average parameter ${\bf b}$.

\subsection{Estimation of the stable parameters}\label{s23}

After estimating the autoregressive parameters, the noise $\{\eps_n\}_{n\in\bbn}$ can be recovered.
Recall that, by $p$-dependence, the noise terms of lags $m>p$ are independent.
Motivated by results for discrete-time stable ARMA processes,
Garcia et al.~\cite{GKM} have applied estimation methods for independent noise variables.
They have also shown in a simulation study that one gets quite reliable estimates by 
treating $R:=\{\eps_n\}_{n\in\bbn}$ as independent sequence. 

By a simple computation we can relate the estimated parameters of the series $R_i$ back to an estimate of the $\alpha$-stable process $L$.
We show this for the CARMA(2,1) model in the next example:

\begin{example}\label{stable2}\rm [Continuation of Example~\ref{stable1}, cf. \cite{GKM}]\\
Using Samorodnitsky and Taqqu \cite{ST}, Property~3.2.2, a relation in distribution between the 
$\alpha$-stable process $L$ and the noise process $\{\eps_n\}_{n \in\bbn}$ of the ARMA(2,1) model 
sampled on a grid with grid size $h$ can be established (for clarification, we will now sometimes write
$(\alpha_L,\gamma_L,\beta_L,\mu_L)$ for the parameters of the $\alpha$-stable process $L$, which were
so far denoted just by $(\alpha,\gamma,\beta,\mu)$). In particular, $\eps_n$ has an $\alpha$-stable 
distribution with parameters $(\alpha_\varepsilon, \gamma_\varepsilon,\beta_\varepsilon,\mu_\varepsilon)$ given by
\begin{align*}
\alpha_\varepsilon &= \alpha_L \, = \, \alpha \\
\gamma_\varepsilon &= \Bigl( \int_0^h \left| \kappa_1 \, e^{\lambda_1 (h-u)} + \kappa_2\, e^{\lambda_2(h -u)} \right|^{\alpha}
 + \left| \kappa_1 \, e^{\lambda_2 h } \,e^{\lambda_1 (h-u)} + \kappa_2\, e^{\lambda_1 h} \,e^{\lambda_2(h -u)} \right|^{\alpha}
 du\Bigr)^{1/\alpha} \gamma_L\\
\beta_\varepsilon  &= \beta_L \frac{\gamma_L^\alpha}{\gamma_\varepsilon^\alpha} \, 
\left( \int_0^h (\kappa_1 \, e^{\lambda_1 (h-u)} + \kappa_2\, e^{\lambda_2(h -u)})^{\langle \alpha \rangle} + 
(\kappa_1 \, e^{\lambda_2 h } \,e^{\lambda_1 (h-u)} + \kappa_2\, e^{\lambda_1 h} \,e^{\lambda_2(h -u)})^{\langle \alpha \rangle} 
\, du \right)\\
\mu_\varepsilon    &= \mu_L \, = \, \mu \quad \text{for }\, \alpha_\varepsilon  \not= 1
\end{align*}
Note that, for $a$ and $p$ being real numbers, $a^{\langle p \rangle} := |a|^p \textnormal{sign} (a)$ 
denotes the signed power (Samorodnitsky and Taqqu \cite{ST}, eq. (2.7.1)). Moreover, we can easily see that 
$\beta_\varepsilon = \beta_L$, if both $\kappa_1$ and $\kappa_2$ are positive.
\end{example}

\subsection{Recovering the states}\label{Lfilter}

In order to calculate the theoretical futures prices derived in Corollary~\ref{forwardprice} it is necessary to recover the states ${\bf X}$ of the CARMA-process.
Brockwell et al.~\cite{BDY} describe a rather ad-hoc method to do this by using an Euler approximation.

In the linear state space model \eqref{eq;statespace}, the Kalman filter is the best linear predictor provided the driving noise is in $L^2$.
Since  $\alpha$-stable L\'evy processes for $\al\in(0,2)$ do not have finite second moments, the Kalman filter will perform unsatisfactorily.
One possibility to resolve this is to apply a particle filter, which does not
require a finite second moment of the noise process. However, the particle filter requires a density function instead, which poses a new problem for $\alpha$-stable processes.
Integral approximations of $\al$-stable densities exist, but they are time consuming to calculate and simple expressions do not exist.
One can use a particle filter by simulating from the $\alpha$-stable distribution, but this is also very time consuming.
A large number of paths need to be simulated in order to get a reasonable estimation (even when using appropriate variance reducing methods like importance sampling).
As an attractive alternative, we introduce a simple $L^1$-filter applicable to CARMA processes with finite mean.

Recall from \eqref{eq;statespace}-\eqref{eq;Y} that we can work with the following state-space representation of the CARMA process
\begin{align}
y_n &= {\bf b}^*\mathbf{x}_n,\\
\mathbf{x}_n &= e^{A h}\mathbf{x}_{n-1} + \mathbf{z}_n\quad \text{with} \quad \,\mathbf{z}_n= \int_{(n-1)h}^{nh} e^{A(nh-u)}\mathbf{e}_p \,dL(u) \label{stateeq}
\end{align}
Here, $y_n$ and $\mathbf{x}_n$ are discrete observations of $Y$ and $\mathbf{X}$, respectively, on a grid with grid size $h$.

Notice that given $y_n$ and $\mathbf{x}_{n-1}$ the value of ${\bf b}^*{\mathbf{z}_n}$ is determined and given by
\begin{align}\label{eq;condz}
{{\bf b}^*\mathbf{z}_n{\bf |}}{y_n,\mathbf{x}_{n-1}} = y_n - {\bf b}^*e^{Ah}\mathbf{x}_{n-1}.
\end{align}
This will come to use in a moment when deriving the filter.

First, we make an "Euler" approximation of the stochastic integral defining $\mathbf{z}_n$ by
$$
\mathbf{z}_n\approx\frac1h\int_{(n-1)h}^{nh}e^{A(nh-u)}\mathbf{e}_p\,du\Delta L(n,h)=-A^{-1}(I-e^{Ah})\frac{\Delta L(n,h)}{h}\,,
$$
where $\Delta L(n,h)=L(nh)-L((n-1)h)$. Note that a traditional Euler approximation (see Kloeden and Platen~\cite{KP}) would use the left end-point value of the integrand in the approximation, whereas here we use the average value of the integrand over the integration interval.  We find
\begin{align}\label{eq;expz}
\mathbb{E}[\mathbf{z}_n \mid y_n,\mathbf{x}_{n-1}] 
& \approx -\mathbb{E}[\Delta L(n,h)/h \mid y_n,\mathbf{x}_{n-1}]\,A^{-1}\,(I - e^{Ah})\, {\bf e}_p\,.
\end{align}
Multiplying \eqref{eq;expz} with ${\bf b}^*$ and combining it with \eqref{eq;condz} gives
\begin{align}\label{eq;condL}
\mathbb{E}[\Delta L(n,h)/h \mid y_n,\mathbf{x}_{n-1}] \approx \frac{y_n - {\bf b}^*e^{Ah}\mathbf{x}_{n-1}}{-{\bf b}^*A^{-1}\,(I - e^{Ah})\, {\bf e}_p}\,.
\end{align}
By plugging \eqref{eq;condL} into \eqref{eq;expz} we find
\begin{align}\label{eq;filterz}
\mathbb{E}[\mathbf{z}_n \mid y_n,\mathbf{x}_{n-1}] &\approx  -A^{-1}\,(I - e^{Ah})\, {\bf e}_p \,  \frac{y_n - {\bf b}^*e^{Ah}\mathbf{x}_{n-1}}{-{\bf b}^*A^{-1}\,(I - e^{Ah})\, {\bf e}_p}.
\end{align}
We can use this as an $L^1$-filter for $\mathbf{z}_n$. Applying \eqref{eq;filterz}, we can filter the states $\mathbf{X}$ of the CARMA-process. Using the state equation \eqref{stateeq} we find
\beam\label{eq;filterx}
\mathbb{E}[\mathbf{x}_n\mid y_n,\mathbf{x}_{n-1}] &\approx& e^{Ah}\mathbf{x}_{n-1} + \mathbb{E}[\mathbf{z}_n \mid y_n,\mathbf{x}_{n-1}].
\eeam

\begin{figure}[ht]
\begin{center}
\includegraphics[width=14cm]{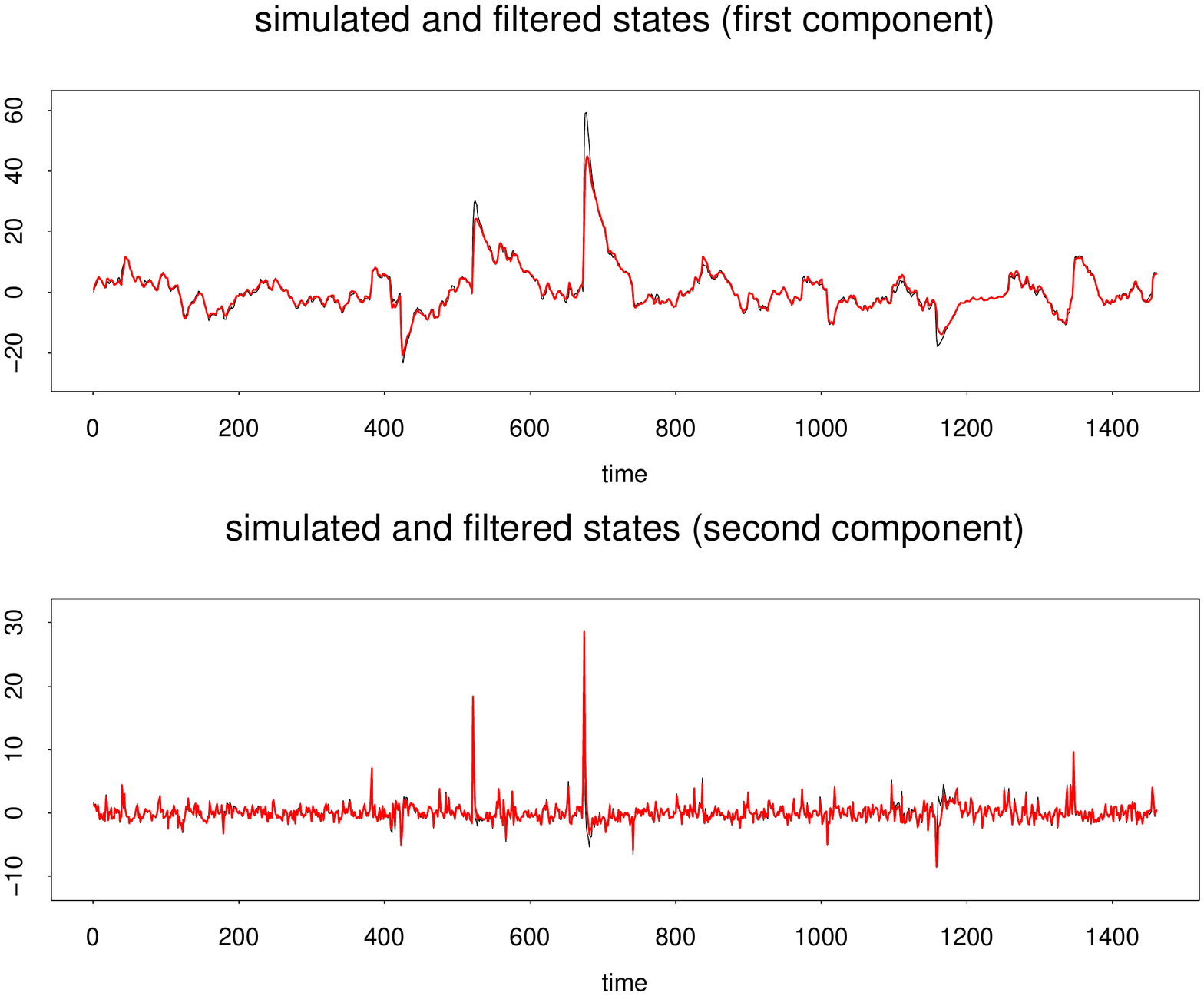}
\end{center}\caption{{\it Estimated states (red) and true states (black) of a simulated $\CARMA(2,1)$ process 
using the $L^1$-filter.}}\label{F;states}
\end{figure}

We tested the filter on simulated data from a CARMA(2,1) process with the same parameters as we find 
from our model for the base spot prices (see Table~\ref{T;est}). The path of the CARMA(2,1) process was 
simulated based on an Euler scheme on a grid size of 0.01 for $0\le t\le 1461$, and the $\alpha$-stable 
L\'evy process was simulated using the algorithm suggested by Chambers, Mallows and Stuck \cite{CMS}. 
The estimation of the states is done on a grid with grid size $h=1$. Figure \ref{F;states} 
shows the estimated states (red curve) for both state components together with the simulated states (black curve). It is 
clearly visible that the $L^1$-filter gives a good approximation of the true states ${\bf X}$ 
driving the $\alpha$-stable CARMA process $Y$.

\subsection{Risk premium comparison}\label{estRisk}

In order to find the optimal threshold $\hat{u}^*$ for filtering out the non-stationary process $Z$ from the futures data, we compare the empirically observed risk premium with its theoretical counterpart.

\begin{figure}[ht]
\begin{center}
\vspace*{0cm}
\includegraphics[width = 14cm]{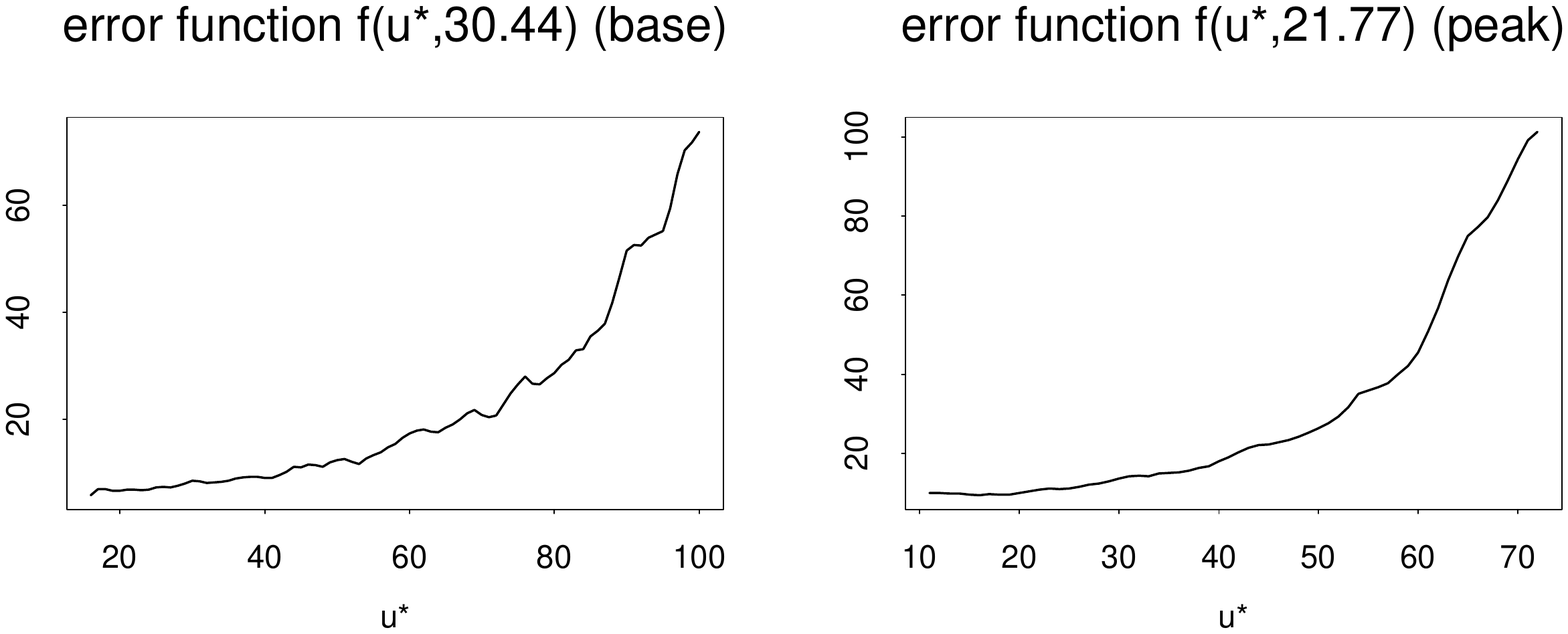}
\vspace*{-5cm}
\caption{{\it The risk premium error function for base data (left) and peak data (right), 
for $v/2 \le u^* \le M_f/2$. In both cases, $f$ has its minimum at $\hat{u}^* = 16$.
}}\label{PLOTfustarBP}
\end{center}
\end{figure}

Recall the risk premium $R_{pr}$ in \eqref{riskpremium} implied by the futures price dynamics in Corollary~\ref{forwardprice}. By using
$v=T_2-T_1$ and recalling the notation $u=\frac12(T_1+T_2)-t$, we can rewrite $R_{pr}$ for $u\ge \frac{1}{2}(T_2-T_1)$ and fixed
$v$ (being one month in our studies) to
\begin{align}\label{riskpremium3}
R_{pr}(u^*,u,v) &= - \frac{1}{v} {\bf b}^*A^{-2} \left( e^{ \frac{1}{2}Av}- e^{- \frac{1}{2}Av} \right) e^{Au} {\bf e}_p \, \left(\mathbb{E}_{Q}[L(1)]-\mathbb{E}[L(1)]\right)  \nonumber\\
&\quad + \textbf{b}^* A^{-1} {\bf e}_p\,\left(\mathbb{E}_{Q}[L(1)]
-\mathbb{E}[L(1)]\right) + u\, \mathbb{E}_Q[Z(1)]\,.
\end{align}
Note that we can estimate all parameters in \eqref{riskpremium3} only depending on a chosen threshold $u^*$. Hence, 
the risk premium in Equation \eqref{riskpremium3} also depends on $u^*$.
In order to find an optimal $\hat{u}^*$ we compare the risk premium \eqref{riskpremium3}, which has been estimated 
on our model assumptions, with the mean empirical risk premium based on the futures prices given by
\begin{align}\label{riskpremium4}
\widetilde{R}_{pr}(u^*,u,v) &:= \frac{1}{v} {\bf b}^*A^{-2} \left( e^{ \frac{1}{2}Av}- e^{- \frac{1}{2}Av} \right) e^{Au} {\bf e}_p \, \mathbb{E}[L(1)] - \textbf{b}^* A^{-1} {\bf e}_p\,\mathbb{E}[L(1)]\nonumber\\
&\qquad+\frac{1}{{\rm card }\, U(u,v)} \sum_{t,T_1,T_2 \in U(u,v)} \Bigl[F(t,T_1,T_2) - \frac{1}{T_2-T_1} \int_{T_1}^{T_2} \Lambda(\tau) d\tau \\
&\qquad- \frac{\textbf{b}^* A^{-1}}{T_2 -  T_1} \left(e^{AT_2} - e^{AT_1} \right) e^{-At} \, \textbf{X}(t)  - Z(t) \Bigr]\nonumber.
\end{align}
 Here,
 \begin{align*}
 U(u,v) := \Big{\lbrace} t,T_1,T_2 \in \mathbb{R} \,{\bf ;} \, \frac{1}{2} (T_2+T_1) -t = u,\, T_2-T_1 = v \text{ and } F(t,T_1,T_2) \text{ exists } \Big{\rbrace}\,.
 \end{align*}
The dependence on the threshold $u^*$ is only implicit.
The estimated sample paths of $Z$ and $Y$ depend on $u^*$, therefore, also the CARMA parameters $A,{\bf b}$, the stable parameters $(\alpha,\beta,\gamma,\mu)$ and the estimated sample paths of the states $\mathbf{X}$ also depend on $u^*$.
In order to compute $\widetilde{R}_{pr}$ and $R_{pr}$ all these estimated parameters are used. 
Consequently, by using different thresholds we will get different estimates and different risk premia.
We want to choose an optimal threshold $\hat{u}^*$, such that the mean empirical  
risk premium $\widetilde{R}_{rp}$ is as close as possible to the model based risk-premium $R_{pr}$.
We invoke a least squares method, i.e. we minimize (for fixed $v$) the mean square error between 
the two functions $(\widetilde{R}_{pr}(u^*,u,v))_{u\ge \frac12 v}$ and $(R_{pr}(u^*,u,v))_{u\ge \frac12 v}$ with respect to all chosen 
thresholds $u^* \in U^*$ for the estimation procedure, cf. Section \ref{est;Lambda}.
\begin{align*}
\hat{u}^*=\text{argmin}_{{u^* \in U^*}} \, \sum_{u = v/2}^{M_{f}} \vert \widetilde{R}_{pr}(u^*,u,v) -R_{pr}(u^*,u,v) \vert^2
\end{align*}
Here the dependence of the error function
\begin{equation}
f(u^*, v) := \sum_{u = v/2}^{M_f} \vert \widetilde{R}_{pr}(u^*,u,v) -R_{pr}(u^*,u,v) \vert^2\, \quad (u^* \in U^*)
\end{equation}
on $u^*$ is only implicit. In our data $v$ corresponds to the average number of days per month 
(i.e. $v=1461/48=30.44$ for the base data and $v=1045/48=21.77$ 
for the peak data), and the number $M_f$ is the longest time to maturity, 
which we recall from Section \ref{est;Lambda} being 200 for the base load contracts and 144 for the peak load.
In order to calculate this minimum we calculate the values of $f(u^*,v)$ for all $u^* \in \left[ v/2, M_f/2 \right] \cap \bbn$.
Figure \ref{PLOTfustarBP} shows the risk premium error function $f(u^*,v)$ for base load (left) and peak load (right). 
In both cases, the minimum is attained at $\hat{u}^* = 16$.  
So our estimation procedure considers only base load forward contracts 
with delivery at least (about) two weeks away, and peak load contracts 
with delivery at least (about) three weeks away. 

Below we present a summary of the estimation algorithm again.
\subsubsection*{The algorithm}

Estimate $\Lambda(\cdot)$ as in \eqref{season-base} for the base load model and as 
in \eqref{season-peak} for peak load model, and subtract from $S(\cdot)$.\\

For each threshold $u^* \in U^*$:
\begin{itemize}
\item Approximate $\mu_{\wt F}(u)= C + u\, \mathbb{E}_{Q}[Z(1)]$ for $u \ge u^*$ and 
estimate $C$, $\mathbb{E}_{Q}[Z(1)]$ by linear regression \eqref{eq;risk};
\item
filter $Z$ by \eqref{eq;z};
\item
model $Y=S-\Lambda-Z$ as CARMA(2,1) process,
estimate the coefficients $a_1,a_2,b_0$ (recall that $b_1=1$) and
estimate the parameters $(\alpha_L,\gamma_L,\beta_L,\mu_L)$ of $L$;
\item
estimate $\mathbb{E}_{Q}[L(1)]$ using \eqref{eq;z};
\item
filter states of $\bX=(X_1,X_2)^*$ using \eqref{eq;filterx};
\item
calculate $R_{pr}(u^*,u,v)$ as in \eqref{riskpremium3} invoking the estimated parameters and states from the former steps;
\item
calculate $\wt R_{pr}(u^*,u,v)$ as in \eqref{riskpremium4} invoking the estimated parameters and states from the former steps 
and the futures data.
\end{itemize}
Now define 
the mean square error of the estimated $R_{pr}(u^*,u,v)$ and $\wt R_{pr}(u^*,u,v)$ based on all different thresholds $u^* \in U^*$.
The optimal threshold is found to be $\hat{u}^*$.

\section{Estimation results}

We now report the other results from the estimation procedure, when using the the optimal threshold $\hat{u}^*=16$
both for base and peak load data. In this section we discuss the estimated values and their implications.

\subsection{Distributional properties of the filtered sample path of $Z$}\label{disZ}

\begin{figure}[htb]
\begin{center}
\vspace*{0cm}
\includegraphics[width = 14cm]{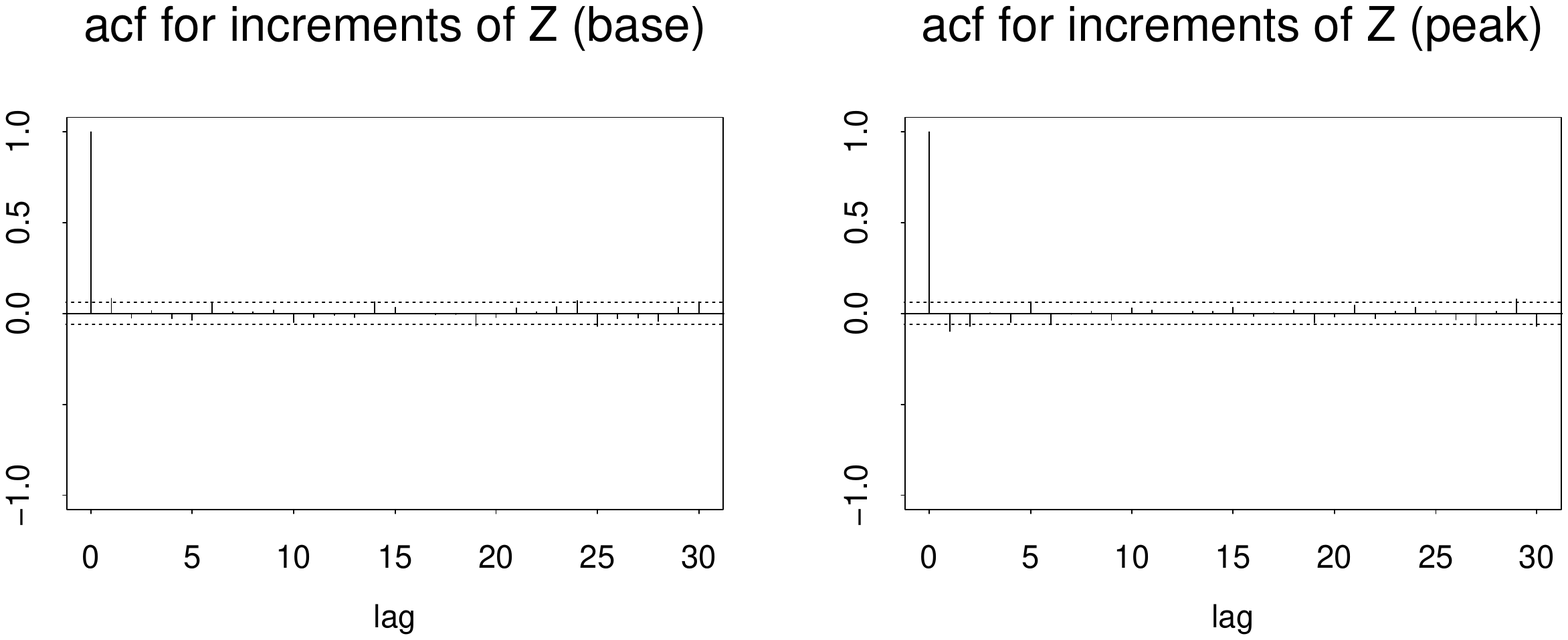}
\vspace*{-5cm}
\caption{{\it Empirical autocorrelation functions for the increments of $Z$, 
for base data (left) and peak data (right).
}}\label{PLOTZacfsBP}
\end{center}
\end{figure}

For the filtered $Z$ which was found using \eqref{eq;z} we can derive certain properties. 
Both for the base and the peak data, the realization of $Z$ shows uncorrelated increments, 
see Figure~\ref{PLOTZacfsBP}. 

\begin{figure}[htb]
\begin{center}
\vspace*{0cm}
\includegraphics[width = 14cm]{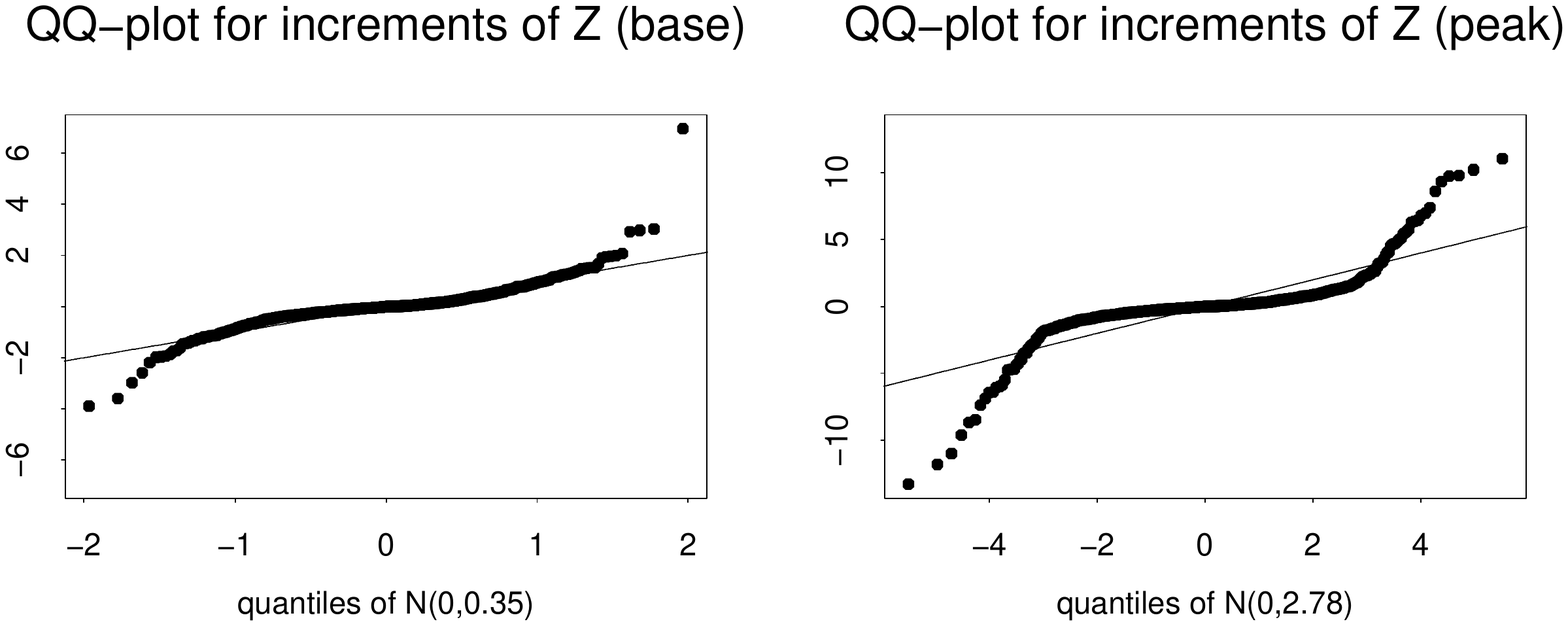}
\vspace*{-5cm}
\caption{{\it QQ-plots for the increments of $Z$ against suitable normal distributions, 
for base data (left) and peak data (right). The 
reference distributions are N(0,0.35) and N(0,2.78) for base and peak data, respectively.
}}\label{PLOTqqBP}
\end{center}
\end{figure}

Figure~\ref{PLOTqqBP} shows QQ-plots for the increments of $Z$ versus a corresponding normal distribution,
both for base data (left) and peak data (right). Note that the empirical variances of the increments
are 0.35 and 2.78 for the base and peak data, respectively. From these plots we conclude that
for both data sets the increments of $Z$ have heavier tails than the Gaussian distribution, and that this feature
is even more pronounced for the peak data. 

\begin{figure}[t]
\begin{center}
\vspace*{0cm}
\includegraphics[width = 14cm]{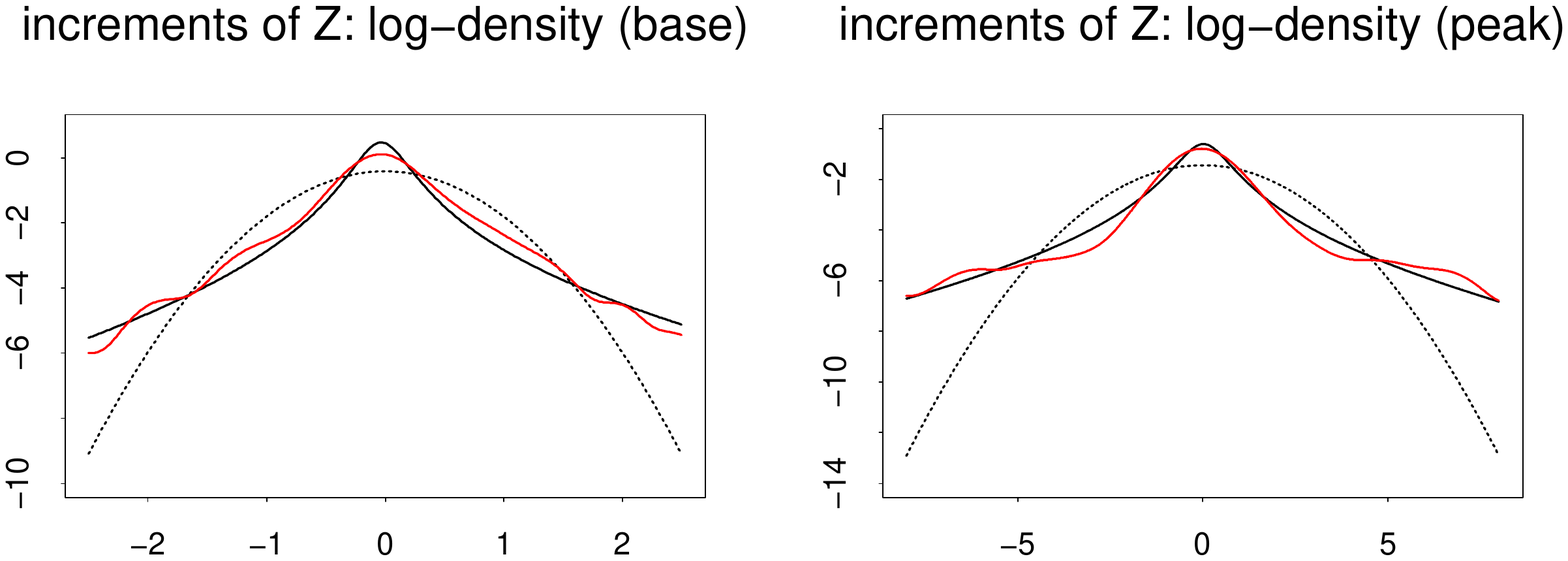}
\vspace*{-5.5cm}
\caption{{\it Log-densities for the incremets of $Z$ (red) as well as for NIG distributions (black solid) 
and normal distributions (black dashed) that have been fitted to these increments, for base data (left) 
and peak data (right).
}}\label{PLOTlogdensBP}
\end{center}
\end{figure}

Kernel density estimates suggest that the increments of $Z$ can be described quite well using 
a normal inverse Gaussian (NIG) distribution (we refer to Barndorff-Nielsen \cite{BN} for a 
thorough discussion on the NIG distribution and its properties). The red curves in 
Figure \ref{PLOTlogdensBP} show log-density estimates for the increments of $Z$ for the 
base data (left) and the peak data (right), respectively. For comparison, we also plot the 
log-density curves of NIG distributions (black solid curves) and normal distributions 
(black dashed curves) that have been fitted to the increments of $Z$ via maximum likelihood 
(the parameters for the NIG distributions can be found in Table~\ref{TableNIG}).
Clearly, the NIG distribution gives a much better fit than the normal distribution. 
Hence, we identify the non-stationary process $Z$ with a normal inverse Gaussian L\'evy process.

\begin{table}
\begin{center}
\begin{tabular}{|l|c|c|c|c|}
\hline
&$\alpha_Z$&$\beta_Z$&$\delta_Z$&$\mu_Z$ \\
\hline
{\bf Base load}& $0.6451$ & $\phantom{-}0.0998$ & $0.2206$ & $-0.0346$ \\
\hline
{\bf Peak load}& $0.2371$ & $-0.0083$           & $0.6582$ & $\phantom{-}0.0230$ \\
\hline
\end{tabular}\caption{{\it Estimated parameters of the NIG distribution $Q$ for the increments of $Z$. 
Since we assume that $\bbe[Z]=0$, the parameters have been estimated conditionally on this assumption.}}
\label{TableNIG}
\end{center}
\end{table}

\brem
Recall that futures contracts are only traded on weekdays and, therefore, no variability 
in $Z$ during weekends is observed. For base load contracts we have thus assumed that $Z$ is constant during weekends 
for filtering purposes, but only considered weekdays data for analysing the distributional properties of $Z$. 
As we already mentioned in Section \ref{estZ}, this strategy could lead to larger up-- or downward movements of $Z$ on Mondays,
when all information accumulated over the weekend is subsumed at once. However, we do not find such a behaviour
in the estimated increments of $Z$. To see this, we calculated the variance of the increments of $Z$ in the base
data set for different weekdays separately. We find a variance of 0.3723 for the increments occurring from Fridays to
Mondays, and a variance of 0.3526 for the remaining increments, i.e.~Mon/Tue, Tue/Wed, Wed/Thu, Thu/Fri. 
For comparison, the corresponding variance from Wednesdays to Thursdays only is 0.3202, whereas for the
remaining increments, i.e.~Mon/Tue, Tue/Wed, Thu/Fri, Fri/Mon, the value is 0.3649.  
Furthermore, the estimated overall variance of the increments of $Z$ is 0.35. 
Taking the estimation error for all these variances into account,
we do not observe a significantly higher variance for the increments from Friday to Monday in the base data.
This supports the idea that no relevant information enters the futures
market over the weekends. Maybe this even can be expected, since the futures deal with a product 
to be delivered {\em quite far} in the future; hence, only {\em really influential} news with a long-range impact 
should affect the market. 

\erem

\subsection{Estimation of the CARMA parameters}

The autocorrelation and partial autocorrelation function of the data suggest that there are two significant 
autoregressive lags, but also a relevant moving average component.
Also the AIC and BIC criterion confirm that a CARMA(2,1) model leads to the best fit.
The estimated parameters of a CARMA(2,1) model are given in Table~\ref{T;est}.
\begin{table}[ht]
\begin{center}
\begin{tabular}{|l|c|c|c|c|c|c|c|}
\hline        
               & \multicolumn{3}{|c|}{CARMA parameters} & \multicolumn{4}{|c|}{Stable parameters} \\
\cline{2-8}
               & $a_1$    & $a_2$    & $b_0$    & $\alpha_L$ & $\beta_L$ & $\gamma_L$ & $\mathbb{E}[L(1)]$ \\
\hline
{\bf Base load}& $1.4854$ & $0.0911$ & $0.2861$ & $1.6524$   & $0.3911$  & $6.4072$   & $\phantom{-}0.0566$ \\
\hline
{\bf Peak load}& $2.3335$ & $0.2263$ & $0.6127$ & $1.3206$   & $0.0652$  & $6.5199$   &           $-0.0448$ \\
\hline
\end{tabular}
\caption{{
\it Estimates of the CARMA parameters and of the parameters of the stable process $L$.
}}\label{T;est}
\end{center}
\end{table}

For the estimated parameters $(\wh a_1,\wh a_2)$ in the autoregression matrix $A$ the eigenvalues of 
$A$ are real and strictly negative, being $\lambda_1=-0.0641, \lambda_2=-1.4213$ for the base load and 
$\lambda_1=-0.1014, \lambda_2=-2.2319$ for the peak load. Our parameters satisfy 
Assumptions \ref{assump}. Hence, the estimated model is stationary.


The estimates of $\alpha_L$ (1.6524 for base and 1.3206 for peak) confirm that 
extreme spikes are more likely in the peak load data. 
As we can conclude from the positive signs of the skewness parameter $\beta_L$, positive 
spikes are more likely to happen than negative spikes for both data sets. 
Note, however, that a direct comparison of the values of $\beta_L$ for the base and the peak
load data is misleading, due to the significantly different parameters $\alpha_L$.
Indeed, if we calculate the empirical skewness for the estimated $\varepsilon_n$ (cf. Section \ref{s23}) 
directly, we get a value of 0.28 for the base load data and 1.59 for the peak load data 
(cf. the comment on $\beta_\varepsilon = \beta_L$ in Example \ref{stable2}; for the base load data 
$(\kappa_1,\kappa_2)=(0.1636,0.8364)$, and for the peak load data $(\kappa_1,\kappa_2)=(0.2400,0.7600)$). 

\subsection{Market price of risk and risk premium}\label{s66}

We next present the results on the risk premium and the parameters for the market price of risk,
based on our statistical analysis of base and peak load contracts with threshold $\hat{u}^*=16$ days
in both cases. Estimates of the relevant parameters are presented in Table \ref{T;riskv}.
\begin{table}
\begin{center}
\begin{tabular}{|l|c|c|c|c|c|}
\hline
               & $\widehat{\mathbb{E}_Q[Z(1)]}$ & $\wh{C}$ & $\widehat{\mathbb{E}_Q[L(1)]}$ & $\widehat\theta_Z$ & $\widehat\theta_L$ \\
\hline
{\bf Base load}& $-0.0243$ & $1.6587$ & $-0.5282$ & $-0.1093$ & $-0.0021$ \\
\hline
{\bf Peak load}& $-0.0382$ & $3.5678$ & $-1.3178$ & $-0.0168$ & $-0.0552$ \\
\hline
\end{tabular}\caption{{\it Estimates of parameters determining the risk.
}}\label{T;riskv}
\end{center}
\end{table}

Recall that $\widehat{\mathbb{E}_Q[Z(1)]}$ and $\hat{C}$ are found from 
regression \eqref{eq;risk}; using the estimates of the CARMA model from Table~\ref{T;est}, 
we derive an estimate of $\mathbb{E}_Q[L(1)]$.
Having both $\mathbb{E}_Q[L(1)]$ and $\mathbb{E}_Q[Z(1)]$ estimated, we can compute 
the parameters in the respective measure transforms of the NIG L\'evy process $Z$ 
and the stable process $L$. For an NIG L\'evy process we use the fact that an 
Esscher transformed NIG($\alpha_Z,\beta_Z,\delta_Z,\mu_Z$) random variable $Z$ 
is again NIG distributed with parameters ($\alpha_Z,\beta_Z + \theta_Z,\delta_Z,\mu_Z$)
(see e.g. Benth et al.~\cite{BBK}, p.~99). Using the mean of an NIG distributed random variable it holds that
\begin{align*}
\mathbb{E}_Q[Z(1)] = \mu_Z + \frac{\delta_Z(\beta_Z+\theta_Z)}{\sqrt{\alpha_Z^2 - (\beta_Z + \theta_Z)^2}},
\end{align*}
where $\theta_Z$ is the market price of risk for $Z$.
Since estimates for the parameters $\alpha_Z$, $\beta_Z$, $\delta_Z$ and $\mu_Z$ are known from
Table~\ref{TableNIG}, we can use the estimate for $\mathbb{E}_Q[Z(1)]$ together with 
the above equality to obtain an estimate of $\theta_Z$, which results here in
$\widehat \theta_Z =-0.1093$ for the base load and in $\widehat \theta_Z =-0.0168$ 
for the peak load data. Since $\theta_Z$ is estimated negative, more emphasis is 
given to the negative jumps and less emphasis to the positive jumps of $Z$ in the risk 
neutral world $Q$.
We see from the estimate on the risk-neutral expectation of $Z$ that the
contribution from the non-stationarity factor of the spot on the overall
risk premium is negative. This is natural from the point of view of the hedging
needs of producers. The non-stationary factor induces a long-term risk, which
is the risk producers want to hedge using futures contracts.

Lucia and Schwartz~\cite{LS} also find a negative market price of risk associated to 
the non-stationary term in their two-factor models, when analysing data from the NordPool 
market. We recall that they propose a two-factor model, where the non-stationary term is a 
drifted Brownian motion. The negative market price of risk appears as a negative risk-neutral 
drift, which corresponds to a contribution to the risk premium similar to our model.
We refer to Benth and Sgarra~\cite{BS} for a theoretical analysis of the Esscher transform in
factor models applied to power markets.

Using the relations $\ga^\al=c_+ + c_-$ and $\beta=(c_+ - c_-)/(c_+ + c_-)$ in the stable parameters as calculated in 
Example~2.3.3 of Samorodnitsky and Taqqu \cite{ST} and plugging in the estimated parameters 
from Table~\ref{T;est}, we find
\begin{align*}
\widehat{c}_+   &=    \frac{1}{2}(1+\beta_L) \gamma_L^{\alpha_L} = 14.9715  \quad\mbox{and}\quad
\widehat{c}_- \, = \, \frac{1}{2}(1-\beta_L) \gamma_L^{\alpha_L} =  6.5532
\end{align*}
for the base load data, and 
\begin{align*}
\widehat{c}_+   &=    \frac{1}{2}(1+\beta_L) \gamma_L^{\alpha_L} =  6.3342  \quad\mbox{and}\quad
\widehat{c}_- \, = \, \frac{1}{2}(1-\beta_L) \gamma_L^{\alpha_L} =  5.5587
\end{align*}
for the peak load data. Then by using \eqref{EQPL} we can derive an estimate
for $\theta_L$ by
\begin{align*}
\wh{\theta}_L = -\left( 
\frac{\wh{\mathbb{E}_Q[L(1)]} - \wh{\mathbb{E}[L(1)]}}{\Gamma(1-\wh{\alpha_L}) 
(\wh c_+ - \wh c_-)} \right)^{\frac{1}{\wh\alpha_L-1}}
\end{align*}
which leads to $\wh{\theta}_L=-0.0021$ for the base lead data 
and to $\wh{\theta}_L=-0.0552$ for the peak load data.
The market price of risk for the CARMA-factor noise $L$ is also negative,
however, unlike the non-stationary factor a negative sign does not
necessarily lead to a negative contribution to the risk premium.
As we already see in the estimate of the constant $C$ in the regression
\eqref{eq;risk}, we get a positive contribution to the risk premium. There
will also be a term involving time to maturity, which will converge to
zero in the long end of the futures curve. This part of the risk
premium may contribute both positively or negatively. The CARMA factor
is thus giving a positive risk premium for contracts, which start
delivering reasonably soon. Since this factor is accounting for the
short term variations, and in particular the spike risk of the spot,
we may view this as a result of consumers and retailers hedging their
price risk and, therefore, accepting to pay a premium for this. This
conclusion is in line with the theoretical considerations of Benth,
Cartea and Kiesel \cite{BCK}, who showed -- using the certainty equivalence principle --
that the presence of jumps in the spot price dynamics will lead to a positive
risk premium in the short end of the futures curve. Bessembinder
and Lemmon~\cite{BeL} explain the existence of a positive premium in the short end of the
futures market by an equilibrium model, where the skewness in spot prices induced by spikes
is a crucial driver.

\begin{figure}[t]
\begin{center}
\vspace*{0cm}
\includegraphics[width = 14cm]{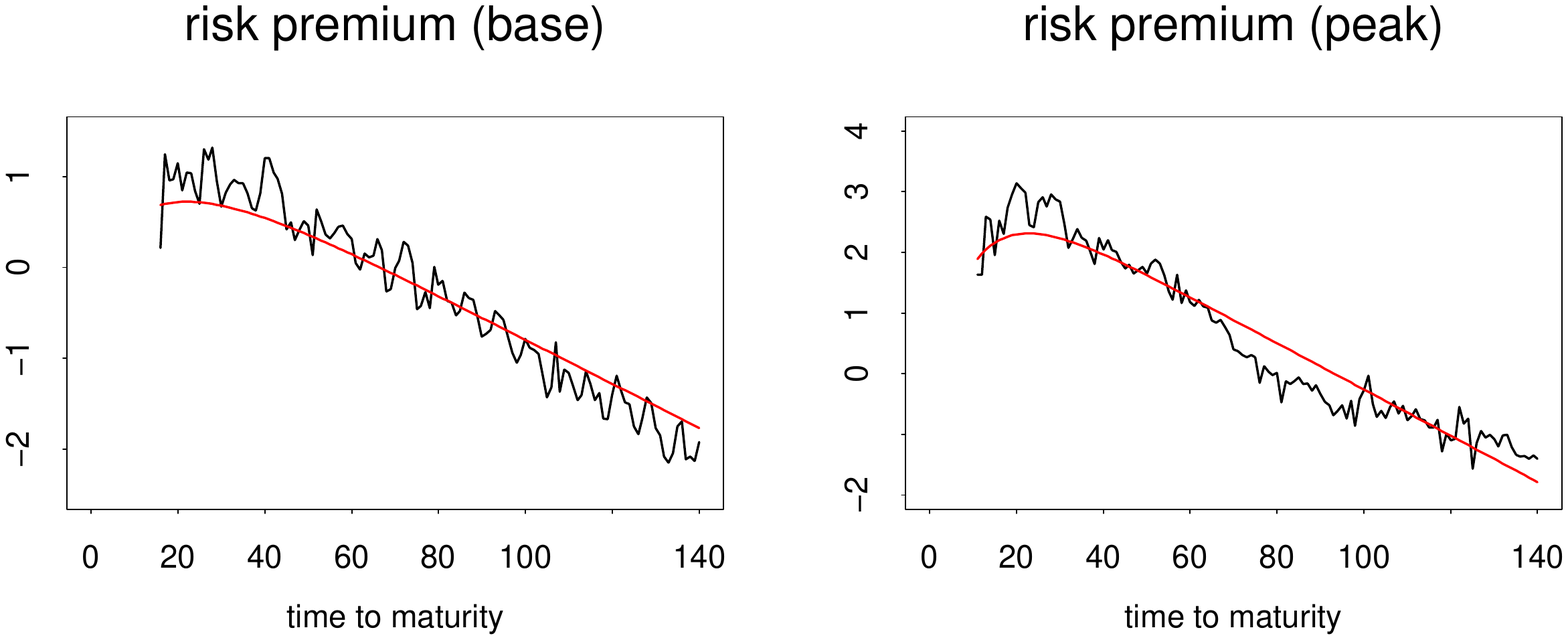}
\vspace*{-5cm}
\caption{{\it The estimated risk premium $R_{pr}$ (red) and the
empirical risk premium $\wt R_{pr}$ (black), for base load data (left) and peak load data (right).
}}\label{F;risk}
\end{center}
\end{figure}

As we know from the third summand in Equation \eqref{riskpremium3}, the risk premium is dominated
by a linear trend for most times to maturity except very short ones. In the latter case,
the first summand of Equation \eqref{riskpremium3} is dominating, leading
to a small exponential decay when time to maturity tends to 0.
A plot of the empirical risk premium versus the theoretical one is given in Figure~\ref{F;risk}.

We see that for the base load contracts the positive risk premium in the short end
of the curve is not that pronounced, however, it is detectable. The risk premium
is negative for contracts starting to deliver in about two months.
On the other hand, the peak load contracts have a clear positive risk premium, 
which changes to a negative one for contracts starting to deliver in
about four months. This form of the risk premium is in line with the 
analysis of Geman and Vasicek \cite{GV}. Interesting here is the difference between 
base and peak load contracts. Base load futures have a longer
delivery period than peak loads, since they are settled against more hours. This means that extreme prices
are more smoothed out for base load contracts. The sensitivity towards spikes are even more pronounced in
peak load contracts, since they concentrate their settlement for the hours, where typically the extreme spikes
occur, and ignore to night hours where prices are usually lower and thus would smooth out the spikes. Hence,
peak load contracts are much more spike sensitive than base load contracts, which we see reflected in the risk premium
having a bigger and more visible positive part in the short end of the futures curve. The study of
Longstaff and Wang~\cite{LW} on the PJM (Pennsylvania, New Jersey and Maryland) market shows that the risk premium may vary over time,
and indeed change sign. Their analysis is performed on hourly prices in the balancing market
as being the spot, and the day ahead hourly prices as being futures contracts. Hence, the analysis
by Longstaff and Wang~\cite{LW} is valid for the very short end of the futures curve.

Our findings for the EEX are in line with the empirical studies in Benth et al.~\cite{BCK}, which applies a
two factor model
to analyse the risk premium in the EEX market. Their model consists of two stationary processes, one for the
short term variations, and another for stationary variations mean-reverting at a slower speed. Their
studies confirm a change in sign of the risk premium as we observe for our model. Moreover, going back to
Lucia and Schwartz~\cite{LS}, they find a positive contribution to the risk premium from their
short-term variation factor, when applying their analysis to NordPool data. This shows that also in this
market there is a tendency towards hedging of spike risk in the short end of the futures curve. On the other
hand, our results for the EEX market are at stake with the findings in Kolos and Ronn~\cite{KR}. They perform
an empirical study of many power markets, where they estimate market prices of risk for
a two-factor Schwartz and Smith model. In the EEX market, they find that both the short and the long term
factors contribute negatively\footnote{In their paper, the signs are positive due to the choice of parametrization of the market price of risk} for the case of the EEX market. However, interestingly,
the PJM market in the US, which is known to have huge price variations with many spikes observed, they
find results similar to ours. We find our results natural in view of the spike risk fully accounted for in
the short term factor, and the natural explanation of the hedging pressure from producers in the non-stationary
factor. Our statistical analysis also strongly suggest non-Gaussian models for both factors, which is very
different to the Gaussian specification of the dynamics in Kolos and Ronn~\cite{KR}.



\section{Conclusion}

In this paper we suggest a two-factor arithmetic spot model to analyse power futures prices.
After removal of seasonality, a non-stationary long term factor is modelled as a L\'evy process, while the short term variations
in the spot price is assumed to follow a stationary stable CARMA process. An empirical analysis
of spot price data from the German power exchange EEX shows that a stable CARMA processes is able to
capture the extreme behavior of electricity spot prices, as well as the more normal variations
when the market is in a quite period.

As in Lucia and Schwartz~\cite{LS} we use a combination of a deterministic function and a non-stationary
term to model the low frequency long term dynamics of the spot. Empirical data suggests that futures
curves and spot prices are driven by a common stochastic trend, and it turns out that this is
very well described by a normal inverse Gaussian L\'evy process. This leads to realistic predictions
of the futures prices. Moreover, a CARMA(2,1) process is statistically the best model for the short term
variations in the spot dynamics.

We apply the Esscher transform to produce a parametric class of market prices of risk for the
non-stationary term. The $\alpha$-stable L\'evy process driving the CARMA-factor is
transformed into a tempered stable process in the risk neutral setting. The spot price
dynamics and the chosen class of risk neutral probabilities allow for analytic pricing of the
futures. A crucial insight in the futures price dynamics is that the stationary CARMA effect from the
spot price is vanishing for contracts far from delivery, where prices essentially behave like the non-stationary
long-term factor.

We propose a statistical method to calibrate the suggested spot and futures model to real data.
The calibration is done using spot and futures data together, where we applied futures prices in the far end
of the market to filter out the non-stationary factor in the spot. We choose a threshold for what is
sufficiently ``far out'' on the futures curve by minimizing the error in matching the
theoretical risk premium to the empirical. In this minimization over thresholds, we need to re-estimate the whole
model until the minimum is attained. Since $\alpha$-stable processes are not in  $L^2$, we introduce a
robust $L^1$-filter in order to recover the states of the CARMA process required for the estimation of the
risk premium.

Our model and calibration technique is used on spot and futures data collected at the EEX. Moreover, in
order to gain full insight into the risk premium structure in this market, we study both peak load and base
load futures contracts with delivery over one month. The base load futures are settled against the hourly
spot price over the whole delivery
period, while the peak load contracts only deliver against the spot price in the peak hours from 8 a.m. to 8 p.m.
on working days. Our model and estimation technique seem to work well in both situations.

We find that the base load futures contracts have a risk premium which is close to linearly decaying with
time to delivery. The risk premium is essentially governed by the long term factor. There is evidence
of a positive premium in the short end of the futures curve. For peak load contracts, which are much more
sensitive to spikes, the positive premium in the short end is far most distinct, but also here the premium
decays close to linearly in the long end of the market. These observations are in line other
theoretical and empricial studies of risk premia in electricity markets, which argue that the risk premia
in power markets are driven by hedging needs. Our findings also show that we should have an exponential
dampening of the premium towards maturity, resulting from the
CARMA factor of the spot.

\subsubsection*{Acknowledgments}
The STABLE package is a collection of algorithms for computing densities,
distribution functions, quantiles, simulating and estimating for $\alpha$-stable distributions.  
A free basic version is available from {\sl academic2.american.edu/$\sim$jpnolan}; 
an advanced version is available from {\sl www.robustanalysis.com}. 
We are grateful to John Nolan for making the packages available.

FEB gratefully acknowledges the financial support from the project "Energy markets: modelling, 
optimization and simulation (EMMOS)",
funded by the Norwegian Research Council under grant 205328/v30. 

CK gratefully acknowledges financial support 
by the Institute for Advanced Study of the Technische Universit\"at M\"unchen (TUM-IAS).

\renewcommand{\baselinestretch}{0.93}

\end{document}